\documentclass[twocolumn, trackchanges]{aastex63}
\usepackage{amsmath}
\usepackage{CJK}
\usepackage{mathrsfs}
\usepackage{xcolor}

\newcommand\hst{3D-{\sl HST}}
\newcommand\prospector{\texttt {Prospector-$\alpha$}}
\newcommand\ML{M/L_\textsl{g}}
\newcommand\stellarML{$M_*/L$}
\newcommand\logML{\log(M/L_\textsl{g})}
\newcommand\gr{$(\textsl{g}-\textit{r})$}
\newcommand\ri{$(\textsl{r}-\textit{i})$}
\newcommand\Mdyn{M_\mathrm{dyn}}
\newcommand\posterior{P(\boldsymbol{\rho}|\boldsymbol{D})}
\newcommand\likelihood{P(\boldsymbol{D}|\boldsymbol{\rho})} 
\newcommand\bD{\boldsymbol{D}}
\newcommand\brho{\boldsymbol{\rho}}
\newcommand\btheta{\boldsymbol{\theta}}

\graphicspath{{./}{figures/}}

\begin{document}
\begin{CJK*}{UTF8}{gbsn}

\title{Flexible Models for Galaxy Star Formation Histories Both Shift and Scramble the Optical Color-M/L Relationship} 

\author[0000-0002-0682-3310]{Yijia Li (李轶佳)}
\email{yzl466@psu.edu}
\affiliation{Department of Astronomy \& Astrophysics, The Pennsylvania State University, University Park, PA 16802, USA}
\affiliation{Institute for Gravitation and the Cosmos, The Pennsylvania State University, University Park, PA 16802, USA}

\author[0000-0001-6755-1315]{Joel Leja}
\affiliation{Department of Astronomy \& Astrophysics, The Pennsylvania State University, University Park, PA 16802, USA}
\affiliation{Institute for Gravitation and the Cosmos, The Pennsylvania State University, University Park, PA 16802, USA}
\affiliation{Institute for Computational \& Data Sciences, The Pennsylvania State University, University Park, PA 16802, USA}

\shorttitle{color-\stellarML{}}
\shortauthors{Li et al.}

\begin{abstract}
The remarkably tight relationship between galaxy optical color and stellar mass-to-light ratio ($M_*/L$) is widely used for efficient stellar mass estimates. However, it remains unclear whether this low scatter comes from a natural order in the galaxy population, or whether it is driven by simple relationships in the models used to describe them. In this work we investigate the origins of the relationship by contrasting the derived relationship from a simple 4-parameter SED model with a more sophisticated 14-dimensional \prospector{} model including nonparametric star formation histories (SFHs). We apply these models to 63,430 galaxies at $0.5<z<3$ and fit a hierarchical Bayesian model (HBM) to the population distribution in the \gr{}--$\logML$ plane. We find that \prospector{} infers systematically higher \stellarML{} by 0.12\,dex, a result of nonparametric SFHs producing older ages, and also systematically redder rest-frame \gr{} by 0.06\,mag owing to the contribution from nebular emission. Surprisingly, the combined effects of the \stellarML{} and \gr{} offsets produce a similar average relationship for the two models, though \prospector{} produces a higher scatter of 0.28\,dex compared to the simple model of 0.12\,dex. 
Also, unlike the simple model, the \prospector{} relationship shows substantial redshift evolution due to stellar aging. These expected and testable effects produce overall older and redder galaxies, though the color--\stellarML{} relationship is measured only at $0.5<z<3$.
Finally, we demonstrate that the HBM produces substantial shrinkage in the individual posteriors of faint galaxies, an important first step toward using the observed galaxy population directly to inform the SED fitting priors.
 
\end{abstract}

\keywords{Galaxy evolution, Galaxy masses, Galaxy colors, Spectral energy distribution, Hierarchical models}

\section{Introduction}
\label{sec:introduction}
The stellar mass of a galaxy ($M_*$) encodes rich information about the formation of the galaxy itself. Stellar mass changes through internal star formation activity and external mergers and galaxy interactions. It is a stable property that allows one to connect together galaxy populations across time when the galaxies themselves have disparate ages, dust contents, sSFRs, and colors. It plays a critical role in our understanding of the evolution of galaxies over cosmic time, as massive galaxies tend to form earlier and quench faster than low-mass galaxies (``downsizing"; \citealt{Cowie1996}; \citealt{massmet}). Also, $M_*$ correlates with many galaxy physical properties such as star formation rate (SFR), metallicity ($Z$), and galaxy size. By combining these scaling relationships, $M_*$ can provide us with a baseline knowledge of galaxy properties. 

Since $M_*$ is not an observable, its determination generally relies on spectral energy distribution (SED) fitting of broadband photometry or spectra (e.g.,  \citealt{Papovich2001, Shapley2001, Pforr2012, Conroy2013, Courteau2014}). In SED fitting, the galaxy spectrum is modeled as an assembly of stellar populations of coeval stars with (typically) homogeneous metallicity. In this composite stellar population, the total stellar mass of the galaxy is the integral of the SFH across time plus the effects of stellar mass loss. The SED-fitting process is complex and most informative when performed with large amounts of data as it involves generating template galaxy SEDs and inferring the model parameters by comparing the model SEDs to the data. 

\begin{figure*}
    \centering
    \includegraphics[width=0.9\textwidth]{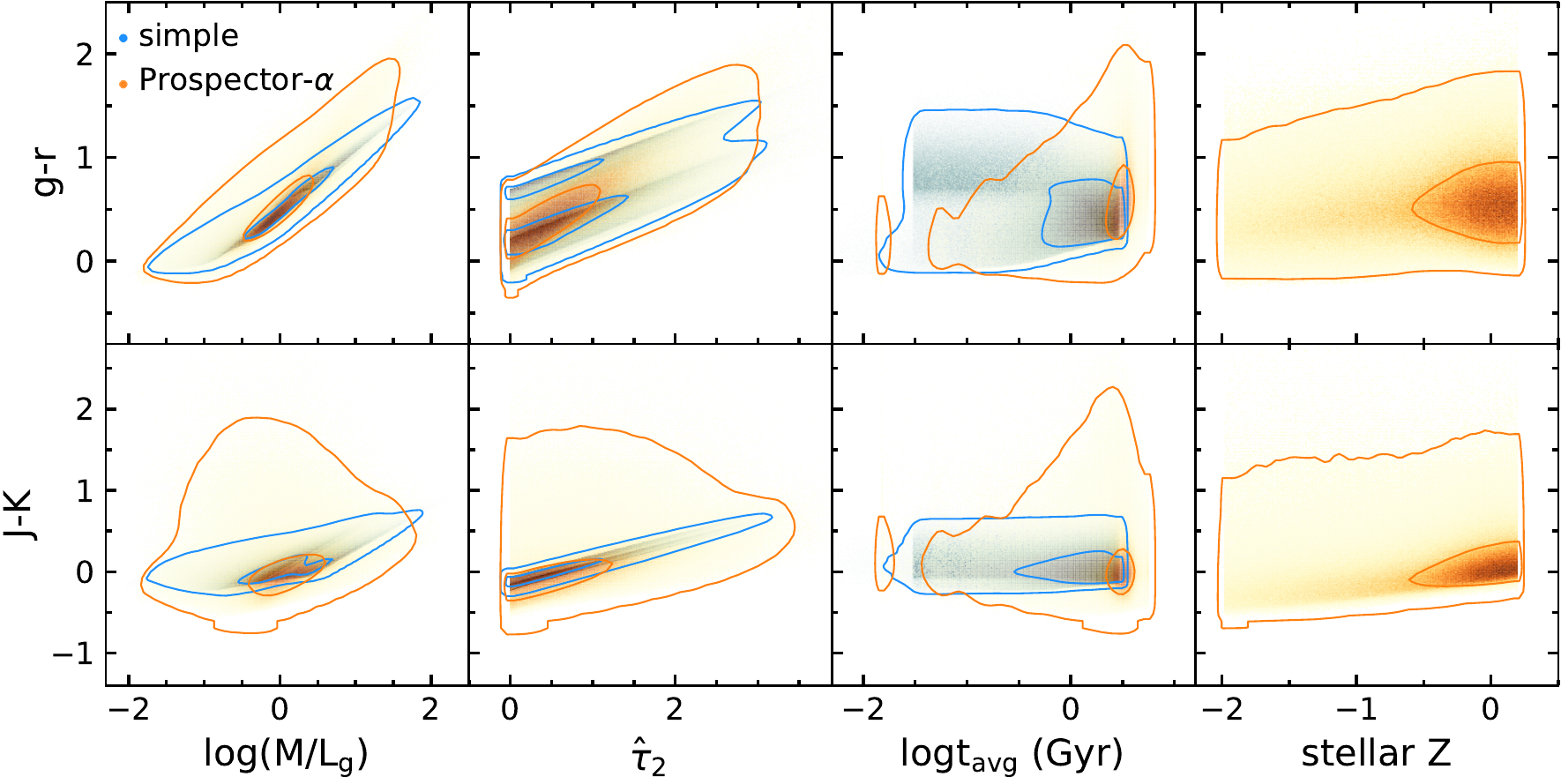}
    \caption{The shaded regions show the relationships between two colors \gr{} and $(\textsl{J}-\textsl{K})$, and four galaxy properties: $\logML$, diffuse dust optical path, age weighted by total mass formed, and stellar metallicity from priors of the simple model (blue) and \prospector{} (orange) at $z=1$. $1-\sigma$ and $3-\sigma$ contours are in solid lines. The SED model priors constrain a notably tight optical color--\stellarML{} relationship. The NIR color cannot constrain \stellarML{} as tightly and no other stellar population parameter is as tightly constrained by a color.} 
    \label{fig:colorpriors}
\end{figure*}

In contrast, when we do not have sufficient data to perform informative SED modeling, one alternative is a popular and efficient method to get $M_*$ using the rest-frame optical color (CMLR for the color--\stellarML{} relationship). This empirical relationship allows remarkably accurate $M_*$ estimates without SED fitting. \citet{Bell&deJong2001} reported a tight linear relationship between the optical color $(B-R)$ and $\log M_*/L$ with a scatter of $\sim$0.2\,dex. Since then several studies have further explored the relationship using combinations of color and $M/L$ in different bands based on both stellar population synthesis (SPS) models and the results of performing SED fitting on the observations (e.g., \citealt{Bell2003}; \citealt{Portinari2004}, \citealt{Zibetti2009}; \citealt{Taylor2011GAMA}; \citealt{Into2013}; \citealt{Courteau2014}; \citealt{McGaugh2014}; \citealt{van-de-Sande2015}; \citealt{Garcia2019}; \citealt{Ge2021}). These studies confirmed the robustness of $M_*$ estimates from a single optical color but revealed a larger scatter in the relationship ($\sim$0.3\,dex). In comparison to optical colors, near-infrared (NIR) colors are less predictive for $M_*/L$, which we will discuss in Figure~\ref{fig:colorpriors} later, and are more sensitive to the modeling of the asymptotic giant branch (AGB) phase of stellar evolution. Because we are able to estimate the approximate color directly from the observations, the optical color--\stellarML{} relationship has been widely used in translating the stellar light to galaxy mass when we know the object redshift, e.g., in dynamical studies \citep[e.g., ][]{Nguyen2020}. 

The notorious dust-age-metallicity degeneracy actually helps shape this remarkably tight relationship. Increasing the age, metallicity, or including more dust will make the galaxy redder and meanwhile enhance the measured $M_*/L$. An important finding of \citet{Bell&deJong2001} is that the dust effect on color and \stellarML{} is parallel to the relationship. These stellar population parameters counteract each other's effects on optical color and \stellarML{} and the net result is a relationship with small scatter. In Figure~\ref{fig:colorpriors} we present the relationships between the priors of color and diffuse dust optical depth ($\hat\tau_2$), mass-weighted age derived from SFH ($t_\textrm{avg}$), and stellar metallicity (stellar $Z$) for the two SED models adopted in this paper\footnote{Note that, only for Figure~\ref{fig:colorpriors}, we match the $\hat\tau_2$ prior of the models to have a fair comparison of the color--\stellarML{} relationship.} (see Section 2.2 for the details and construction of Figure~\ref{fig:colorpriors}). As we have emphasized, a tight relation between optical color and $\logML$ is a fundamental outcome of the stellar physics and dust attenuation models. 

Figure~\ref{fig:colorpriors} shows that due to the strong degeneracies among parameters, we cannot infer the age, dust, and metallicity from the optical color \gr{} alone, as indicated by their broad distribution at a given color. However, we can predict the $\ML$ ratio with high confidence from \gr{}. On the contrary, $(\textsl{J}-\textsl{K})$ color spans a narrow range and is not sensitive to either $\ML$ or stellar population parameters. This is because the degeneracy works differently for different bands. The shape of the degeneracy makes optical \stellarML{} increase much faster than NIR color as the stellar population parameters vary. 

Many components of SED modeling can potentially affect the accuracy of the mass determination from the empirical optical color--\stellarML{} relationship. Such components include the physical model, the initial mass function (IMF), the stellar isochrones, and the stellar spectral libraries. IMF variations can shift the entire relationship to higher or lower \stellarML{} but generally do not influence the variance, as most IMF assumptions only affect the low-mass stars, which emit a small fraction of the total light. 

After choosing the same IMF, different choices of SPS models can still introduce a $\lesssim$\,0.2\,dex offset in \stellarML{} as argued by several previous studies (e.g., \citealt{Zibetti2009}; \citealt{van-de-Sande2015}; \citealt{Ge2019}). The offset is strongest for optically blue stellar populations. There are many possible reasons for the offset such as different treatment of the AGB phase of stellar evolution \citep{Zibetti2009, Kriek2010}, and the completeness of the covered parameter space of the stellar spectral libraries, particularly for young, metal-poor populations \citep{MIST1, Senchyna2019}. However, the influence on the color--\stellarML{} relationship due to SPS models is not the topic of this paper.

In this work, we want to evaluate the role of the simplified prescriptions in SED models which comparatively has not seen as much attention in the literature. 
Previous work on the color--\stellarML{} relationship mostly relies on relatively simple model assumptions like parametric SFHs, fixed dust attenuation curves, and often assumes fixed metallicity. These simple prescriptions have been used for a long time because they make the model straightforward to fit and easy to interpret. Nevertheless, SED-fitting models have become substantially more complex over these years (see reviews by \citealt{Walcher2011}; \citealt{Conroy2013}) in accounting for extra important effects, such as freedom in SFHs, metallicity evolution, nebula emission, dust emission, and AGN, in concert with the development of machinery that creates self-consistent population evolution. It is unclear if the color--\stellarML{} relationship from previous simpler SED models emerges from a substantially more sophisticated framework.%
In this paper, we employ a sophisticated physical model \prospector{} \citep{Prospector-1}, which allows a wide range of physics and has 14 free parameters. By comparing it with a simpler model, we will demonstrate that how we model the galaxy has a large impact on the resultant relationship.

Instead of performing a linear regression to the optical color and $\log M_*/L$ best-fit data like most previous studies, we utilize a hierarchical Bayesian model (HBM) to derive the density distribution of the relationship. Our HBM characterizes the relationship with a population distribution that models the distribution parameter of $\log M_*/L$ at given color with explicit parameters. The hyperparameters that define the population model are fit upon the individual color and \stellarML{} posteriors from SED fittings. The result of this two-level Bayesian inference is the full posterior distribution of the population parameters. 
This method naturally corrects for the observational uncertainties of color and $M_*/L$ by utilizing the SED-fitting posteriors as weights, without any requirement for assuming Gaussian or uncorrelated posteriors. Traditional $\chi^2$ minimizing can be biased by an uneven data distribution in the measurement plane, especially when the measurement uncertainties of the independent variable are comparable to its intrinsic population scatter \citep{Kelly2007}. HBM is likely to be less biased as it naturally distinguishes intrinsic scatter from the measurement errors in its hierarchical structure. Additionally, we pass likelihoods to the population model instead of posteriors by including priors in the weights. In this way we go from an informative prior to an ``uninformative” (flat) prior, which is normally not the case in linear regression.
Also, because HBM assumes the fit objects coming from the same population and assigns a shared prior distribution (i.e., the population distribution) to the individual color and $M_*/L$ estimates, it shrinks the individual fits and makes them cluster around the population mean \citep{Loredo&Hendry2019}. This alludes to one big advantage of the HBM, that we can learn new priors and reapply them to SED fits. 

In this work, we will investigate the relationship between optical color \gr{} and \stellarML{} in the $\textsl{g}$-band $\ML$ using two contrasting SED models. Our goals are to (1) diagnose how SED model assumptions affect the relationship since the \prospector{} model is very different from previous models; (2) derive a relationship at higher redshifts than previous studies, which allows data-based \stellarML{} estimates when the full machinery of SED fitting is impractical. The paper is organized as follows. In Section \ref{sec:sample}, we review the sample properties and the key features of our SED models. In Section \ref{sec:model}, we introduce the algorithms of our hierarchical Bayesian modeling approach. We present the resultant relationships in Section \ref{sec:result}, where we also investigate the driving factor for the \stellarML{} and the color offsets between the two SED models. We compare our results to a few previous works and discuss what we learn from the HBM in Section \ref{sec:discussion}. In Appendix \ref{sec:appendixA} we present a mock test of our HBM. We assume a \citet{Chabrier2003} IMF in our analysis. All colors and \stellarML{} are in the rest-frame band. We use AB magnitudes throughout the paper and adopt the absolute magnitude of the Sun in $\textsl{g}$-band $M_\textsl{g} = 5.11$ \citep{Willmer2018solarMags}.

\section{Data from SED fitting}
\label{sec:sample}
In this work, we will contrast the optical color--\stellarML{} relations derived from two SED models. Both models are constructed within the SED-fitting framework \texttt{Prospector} (\citealt{Prospector-1}; \citealt{Prospector-2}). 
We select our sample from the \hst{} photometric catalogs \citep{Skelton2014}. In Section 2.1 we introduce the photometry from the \hst{} survey and our selection criteria. In Section 2.2 we describe the two SED models used in this work. In Section 2.3 we show the prior distribution of \gr{} and $\logML$, which will be used in deriving the hierarchical models.
\subsection{\hst{} Sample}
\hst{} \citep{Skelton2014} is a space-based grism survey covering $\sim$900\,arcmin$^2$ in five well-studied extragalactic fields. It provides between 17 (the UDS field) and 44 (the COSMOS field) bands of photometry at wavelengths 0.3--8 $\mu$m for hundreds of thousands of galaxies. The survey is supplemented with Spitzer/MIPS 24 $\mu$m photometry from \citet{2014ApJ...795..104W}. The \hst{} catalogs also provide photometric redshifts from the EAZY SED-fitting code \citep{2008ApJ...686.1503B}. Approximately 30\% of galaxies studied in this work have measured spectroscopic redshifts or grism redshifts, which are computed by fitting the photometry and spectrum simultaneously \citep{2016ApJS..225...27M}. 

In this paper, we adopt a subsample of \hst{} galaxies from \citet{2019ApJ...877..140L} and \citet{prospector-massfunc}, consisting of 63,430 galaxies. This sample is selected between $0.5<z<3.0$ above the stellar mass completeness limit of \hst{}, which is the mass of the least-massive galaxy detectable. It is critical when performing population-level inference to reduce or eliminate selection effects by working with a mass-complete sample, or alternatively to model the selection function very well.
With the redshift cut, the observed photometry covers the rest-frame \gr{} color across the full redshift range. Details of the selection criteria and adjustments to the photometric zero-points from the default \hst{} catalogs are described in \citet{2019ApJ...877..140L}.

\subsection{Two Contrasting Physical Models for SED fitting}
We use the \texttt{Prospector} galaxy SED-fitting code to fit the \hst{} photometry. The \texttt {Prospector} inference framework is based on Bayesian forward modeling. The posterior parameter distribution is sampled using the dynamic nested sampling code \texttt{dynesty} \citep{dynesty}. For every galaxy in our sample, we fit a simple SED model that mimics the FAST settings \citep{2009ApJ...700..221K} as used to derive the stellar population parameters in \hst{} catalog \citep{Skelton2014}, and a more complex SED model \texttt {Prospector-$\alpha$} \citep{Prospector-1} to the photometry. The \prospector{} fits have been performed in \citet{2019ApJ...877..140L} and \citet{prospector-massfunc}. 
We adopt MESA Isochrones and Stellar Tracks (MIST; \citealt{MIST0}; \citealt{MIST1}), and MILES stellar spectral libraries \citep{MILES} in the Flexible Stellar Population Synthesis (FSPS; \citealt{Conroy2009}) framework\footnote{A small source of uncertainty in the SED-fitting procedure comes from the interpolation among stellar metallicity grids in simple stellar population (SSP) models \citep{Mitchell2013}, since $Z$ is a complex function of stellar population properties and observed flux. Our color and \stellarML{} are sensitive at the $\sim$0.03\,mag and $\sim$0.03\,dex level to the metallicity interpolation scheme (i.e., triangular weighting versus a delta function).
}.

The simple model is constructed using basic assumptions that have been widely applied in SED modeling. It has four parameters: the stellar mass formed $M_{*, \mathrm{formed}}$, the diffuse dust optical depth $\hat\tau_2$, the galaxy age $t_\mathrm{age}$, and the star formation timescale $\tau$. The stellar metallicity is fixed to solar metallicity. We adopt the \citet{Calzetti2000} dust attenuation curve with a flat prior over $0 < \hat\tau_2 < 4$. The $\hat\tau_2$ parameter controls the normalization of the attenuation curve. We assume an exponentially declining SFH with a minimum $t_\mathrm{f}$ of 30\,Myr.

The 14-parameter \prospector{} model incorporates many of the recent important improvements in SED models into a single consistent framework. Free stellar and gas-phase metallicity are allowed in the model. The stellar mass--stellar metallicity relationship modified from \citet{massmet} is implemented as a prior. We assume a flexible two-component dust attenuation model accounting for birth-cloud and diffuse dust separately. For the diffuse dust component, we add a parameter dust index to enable variations in the shape of the attenuation curve as in \citet{Noll2009}, and include the UV dust absorption bump. Dust emission from energy balance is also built in the model. We use a step function nonparametric SFH with seven time bins, which is more capable of capturing the diversity of galaxy SFHs than the exponentially declining SFH \citep{Leja2019a, Carnall2019, Lower2020}. The model includes both the nebular line and continuum emission as implemented by \citet{Byler2017}, which are important for young stellar populations or high-redshift objects. Mid-infrared dust emission from a dust-enshrouded AGN is also permitted in the model.
The priors of \prospector{} model parameters are demonstrated in \citet{2019ApJ...877..140L}.

\subsection{Priors on Color and Mass-to-light Ratios}
\begin{figure}
    \centering
    \includegraphics[width=.9\columnwidth]{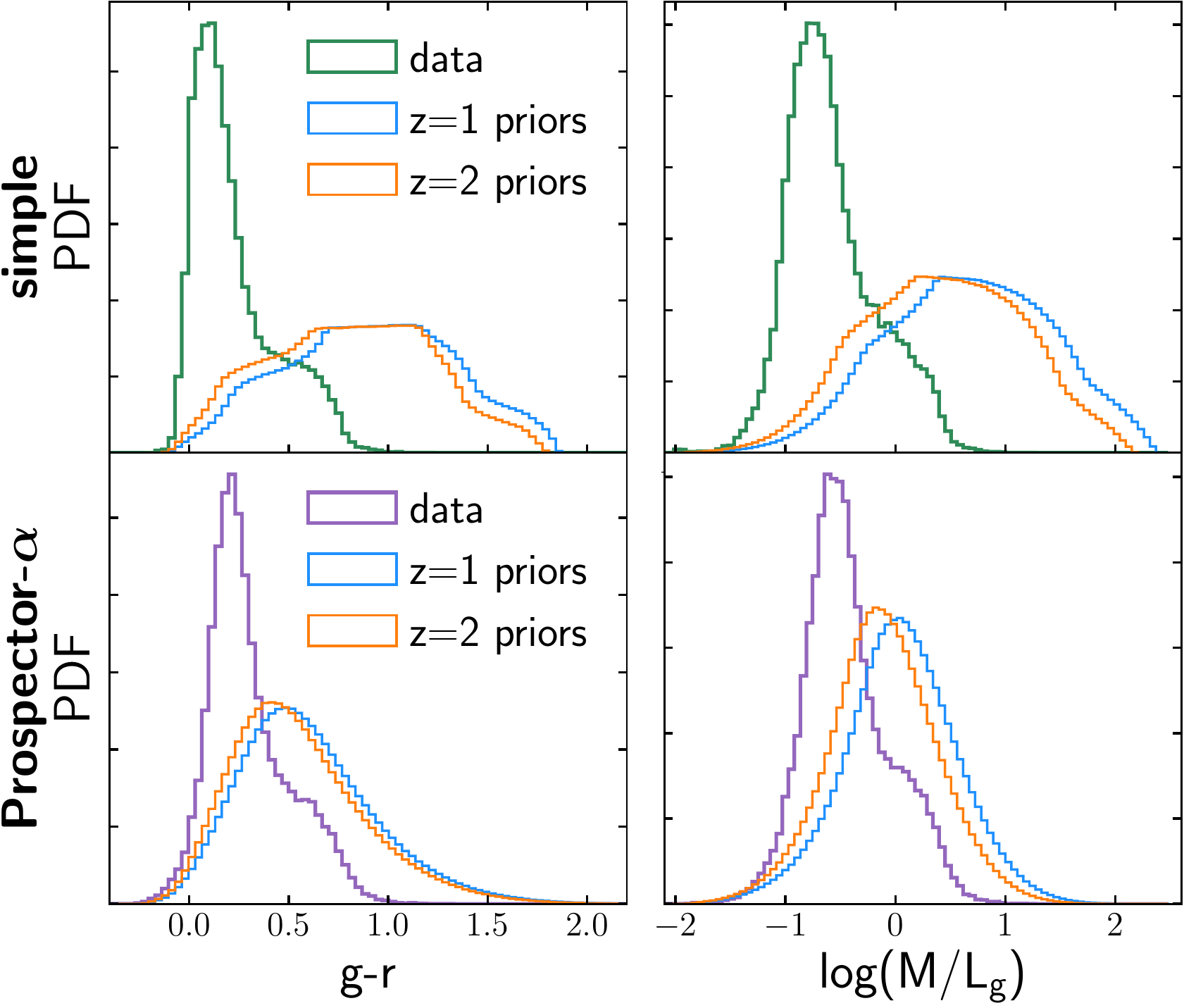}
    \caption{Prior and data distributions for \gr{} and $\logML$. The upper and lower panels show the simple model and \prospector{} respectively. \prospector{} priors describe the data better as the simple priors are too wide. The \prospector{} fits provide systematically higher $\logML$ by 0.12\,dex and higher \gr{} by 0.06\,mag. We show in Section~\ref{sec:deltaML} and Section~\ref{sec:deltagr} that the offsets are mainly driven by different mass-weighted age measurements and the inclusion of nebular emission, respectively.}
    \label{fig:priors}
\end{figure}

To fit a hierarchical model to the optical color--\stellarML{} relationship, we must first infer the SED-fitting priors on both color and $M_*/L$, so we can correct for the priors in the HBM and not be biased by assumptions used in our physical models. Color and $M_*/L$ priors are not specified explicitly, but instead are specified implicitly by the choice of priors on SED model parameters including SFH, dust, metallicity, etc. We infer them numerically, and their joint distributions are shown in Figure~\ref{fig:colorpriors}. Their marginal distributions are shown in Figure~\ref{fig:priors}. The resulting prior probability density distributions will be used in the next section for building hierarchical models. 

We compare the \gr{} and $\logML$ priors and their distribution inferred for the 63,430 galaxies fit in Figure~\ref{fig:priors}. The priors are not closely matched to the data, especially for the simple model. This is because the observed number density of galaxies is dominated by blue galaxies with small $M_*/L$ and the number densities of various subpopulations of galaxies are not yet typically taken into account in galaxy SED-fitting priors. 

Figure~\ref{fig:priors} shows that the \prospector{} priors describe the data better than the simple model, with a narrower range and a peak closer to the data. This is mostly owing to the different dust priors adopted, since galaxies with a high $\hat\tau_2$ will be redder and correspond to higher mass-to-light ratios. The simple model assumes a uniform $\hat\tau_2$ prior from 0 to 4, following the standard for the field (e.g., \citealt{daCunha2008}, \citealt{Marchesini2009}, \citealt{Muzzin2013ApJS}, \citealt{Skelton2014}). Whereas the \prospector{} model has a more informative $\hat\tau_2$ prior, a truncated Gaussian prior with a mean of 0.3 and a standard deviation of 1 between 0 and 4. Consequently, the priors of the simple model typically own more dust content and are redder.

We further demonstrate the redshift dependence of the priors in Figure~\ref{fig:priors}. As the redshift decreases, the prior distributions shift to the redder regime. Priors at different redshifts have the same blue end, implying young galaxies in the nearby universe. Overall, the redshift effects on the \gr{} and $\logML$ priors are not strong. However, we will show later that the observed redshift dependence of the color--\stellarML{} relationship is strong.

\section{A hierarchical model for the color--\stellarML{} relationship} \label{sec:model}
In this section, we will construct a hierarchical Bayesian model to fit the relation between the rest-frame \gr{} color and the stellar mass-to-light ratio using the data described in Section \ref{sec:sample}. Since this work aims to understand the dependence of \stellarML{} on optical colors and the intrinsic scatter in this relation, in the HBM we parameterize the distribution of $\logML$ at a given \gr{} with hyperparameters $\brho$. The model computes posteriors on the population parameters $\brho$ from the individual galaxy fits $\btheta$, where $\btheta = \{\logML_i, \gr{}_i\}$. Our approach is similar to \citet{prospector-massfunc} in the context of modeling the stellar mass function (see also \citet{Nagaraj2022} for HBMs on the relationship between dust attenuation and galaxy properties). We employ the dynamic nested sampling package \texttt{dynesty} \citep{dynesty} to sample the model posteriors.

We model the probability density function (PDF) of $\logML$ at a fixed \gr{} and redshift $z$ with a skewed generalized Student's t (sgt) distribution:
\begin{flalign}\label{eq:ProbML}
P(\btheta_i|\brho) &= \frac{p}{2 \sigma q^{1 / p} \beta\left(\frac{1}{p}, q\right)} \\\nonumber &\left(\frac{|\logML_i-\mu|^p}{q \sigma^p(\lambda \operatorname{sgn}(\logML_i-\mu)+1)^p}+1\right)^{-(\frac{1}{p}+q)}, \nonumber
\end{flalign}
where $\beta$ denotes the beta function and $\operatorname{sgn}$ is the sign function defined by:
\begin{equation}
    \operatorname{sgn} (x) = \begin{cases}-1 & \text { if } x<0, \\ 0 & \text { if } x=0, \\ 1 & \text { if } x>0.\end{cases}
\end{equation}
This generalized Student's t-distribution has 5 parameters. $\mu$ is the mode of the distribution. $\lambda$ controls the skewness of the distribution. $p$ and $q$ are the kurtosis parameters. $\sigma$ accounts for the variance of the distribution.
We model the evolution of $\mu$, $\lambda$, $p$, and $q$ with quadratics in color, and, motivated by the redshift dependence in Figure~\ref{fig:priors}, we additionally include a linear dependence on redshift for the location parameter $\mu$, which accounts for potential variations in the slope of the \gr{}--$\logML$ relation over different redshifts (e.g., \citealt{Szomoru2013}). 
\begin{equation}\label{eq:Parametrization}
\begin{aligned}
&\mu = a_0 + a_1 (\textsl{g}-r)_i + a_2 (\textsl{g}-r)_i^2 + a_3 z, \\ 
&\lambda = b_0 + b_1 (\textsl{g}-r)_i + b_2 (\textsl{g}-r)_i^2, \\
&p = c_0 + c_1 (\textsl{g}-r)_i + c_2 (\textsl{g}-r)_i^2, \\
&q = d_0 + d_1 (\textsl{g}-r)_i + d_2 (\textsl{g}-r)_i^2.&
\end{aligned}
\end{equation}
Here $a_{0,1,2,3}$, $b_{0,1,2}$, $c_{0,1,2}$ and $d_{0,1,2}$ are the quadratic coefficients.
The model has 14 degrees of freedom to control the population-wide behaviors of $\logML$ and \gr{}. Such flexible formalism is necessary for fitting the skewness and the heavy tails of the $\logML$ density distribution. 
Due to the flexibility in the skew and kurtosis parameters, we can achieve a good fit with a fixed value of the variance parameter $\sigma$. This is also motivated by the shape of the observed distribution, where the scatter does not change significantly at different colors. 

The sgt distribution requires $-1<\lambda<1$, $p>0$, and $q>0$. So in practice, we reparametrize the quadratic coefficients $b_{0,1,2}$, $c_{0,1,2}$ and $d_{0,1,2}$ using the values of $\lambda$, $p$, $q$ at ($\textsl{g}-\textit{r}) = -0.3, 0.5, 1.3$, respectively. The reparameterization process is similar to Appendix B of \citet{prospector-massfunc}. Using the anchor points such as $\lambda_{-0.3}, \lambda_{0.5}, \lambda_{1.3}$ instead of $b_{0,1,2}$ makes it easier to set the priors. We are able to satisfy the upper and lower limits of the sgt parameters directly by defining the proper prior range. The anchor points are selected to roughly cover the \gr{} range, and the choice of the specific anchor points should not alter the results.

To summarize, our HBM model has 14 parameters to describe the distribution of the \gr{}--$\logML$ relationship.
We assume uniform priors for these hyperparameters:
\begin{equation}
\begin{aligned}
a_0, a_1, a_2, a_3 &\sim \mathrm{Uniform}(-10, 10), \\ 
\lambda_{-0.3}, \lambda_{0.5}, \lambda_{1.3} &\sim \mathrm{Uniform}(-1, 1), \\
p_{-0.3}, p_{0.5}, p_{1.3} &\sim \mathrm{Uniform}(0, 200), \\
q_{-0.3}, q_{0.5}, q_{1.3} &\sim \mathrm{Uniform}(0, 1500),\\
\sigma &\sim \mathrm{Uniform}(0, 20),.
\end{aligned}
\end{equation}
In Appendix \ref{sec:appendixA} we generate mock data from this population model and validate that we can recover the distribution.

Now that we have defined our model for the distribution of \stellarML{} at a fixed color, we describe how to write down the likelihood for our HBM. The HBM is a straightforward extension of standard Bayesian analysis. Here, however, the input data are the posteriors from the individual galaxy fits, and the output is a model for the population distribution in \gr{} and $\logML$.
By Bayes' theorem, 
\begin{equation}
\posterior = \frac{ \likelihood P(\boldsymbol{\rho})} {P(\boldsymbol{D})}
\end{equation}
Here, $\bD$ is the data vector, and $P(\bD)$ is a normalizing constant that can be ignored in our posterior sampling. $P(\brho)$ is the prior distribution. 
Let $\btheta_i$ represent the $(\logML_i, \gr{}_i)$ vector for galaxy $i$. We can therefore rewrite the likelihood $\likelihood$ in terms of $\btheta_i$:
\begin{equation}
\likelihood = \int d^N \btheta P(\bD | \left\{\btheta_1, \ldots, \btheta_N\right\}) P(\left\{\btheta_1, \ldots, \btheta_N\right\} | \brho),
\end{equation}
where $N$ is the total number of galaxies in our data.
Because the fits are performed independently to each galaxy, 
\begin{equation}\label{eq:LikeIntegral}
\begin{aligned}
\likelihood &= \int d^N \btheta \prod_{i=1}^N P(\bD_i | \btheta_i) P(\btheta_i | \brho) \\
&= \prod_{i=1}^N \int d\btheta_i P(\bD_i | \btheta_i) P(\btheta_i | \brho),
\end{aligned}
\end{equation}
where $P(\btheta_i | \brho)$ is the probability density of $\logML_i$ at $\gr{}_i$. Here our model probability density is weighted by the likelihood $P(\bD_i | \btheta_i)$ from the SED fits performed by \texttt {Prospector}. The model likelihood for each galaxy $P(\bD_i | \brho)$ represents the weighted average of the model density distribution over all possible values of $\theta_i$.

In order to calculate the integral $P(\bD_i | \brho) = \int P(\bD_i | \btheta_i) P(\btheta_i | \brho) d\btheta_i$ from Equation~\ref{eq:LikeIntegral}, we estimate the likelihood $P(\bD_i | \btheta_i)$ from the posteriors and priors we compute during the SED fits in Section \ref{sec:sample}.
We draw $m$ samples $\left\{\btheta_{i, 1}, \ldots, \btheta_{i, m}\right\}$ from the posterior $P(\btheta_i | \bD_i)$ of every galaxy. We choose $m=50$ samples. Such a sample size is enough to resolve the posteriors while also allowing the fit to remain computationally tractable. We assign each posterior sample an importance weight:
\begin{equation}
w_{i, j}=\frac{1}{P(\btheta_{i, j})},    
\end{equation} 
where $P(\btheta_{i, j}$) is the marginalized prior on $\logML_{i,j}$ demonstrated in Figure~\ref{fig:priors}. Accordingly, the model likelihood for each galaxy can be expressed in terms of $P(\btheta_{i, j} | \boldsymbol{\rho})$ from Equation~\ref{eq:ProbML} and $w_{i, j}$: 
\begin{equation}
P(\bD_i | \brho) \approx \frac{\sum_{j=1}^m w_{i, j} P\left(\btheta_{i, j} | \brho\right)} {\sum_{j=1}^{m} w_{i, j}}.
\end{equation}
Therefore, our full log-likelihood becomes 
\begin{equation}
\ln \likelihood 
\approx  \sum_{i=1}^N \ln \left(\frac{\sum_{j=1}^m w_{i, j} P(\btheta_{i, j} | \brho)} {\sum_{j=1}^{m} w_{i, j}}\right).
\end{equation}

\section{Model results}
\label{sec:result}
While Figure~\ref{fig:colorpriors} shows that our model priors predict a tight relation between the optical color and $M_*/L$, in this section we will use real data to determine whether the observed relationship is similarly tight. 
We will first compare the \gr{}--$\logML$ relationship derived from the simple model and from the sophisticated \prospector{} model as described in Section \ref{sec:sample}. We will then explore the origin of their \stellarML{} differences and color differences. In the end, we will apply the new color--\stellarML{} relationship from \prospector{} HBM fit as a prior and show new color and \stellarML{} estimates.

\subsection{Relation Between \gr{} and the Stellar $M/L$ Ratios}
\label{sec:HBMrelation}
\begin{figure*}
    \centering
    \includegraphics[width=\textwidth]{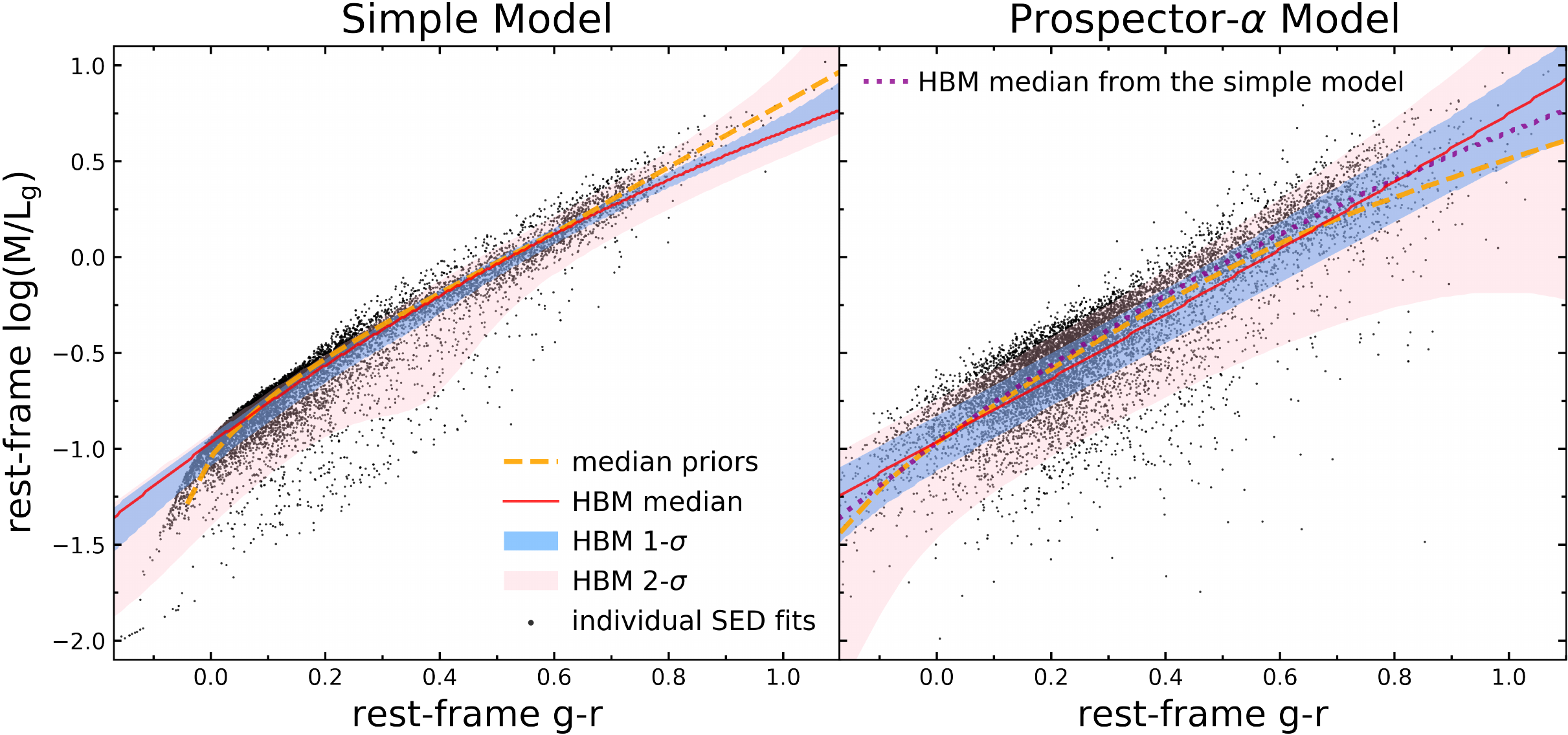}
    \caption{Relationship between the rest-frame \gr{} color and stellar mass-to-light ratio $\logML$ for the simple and \prospector{} model. The yellow dotted lines show the median of the prior distribution. Black dots represent the data from SED fits. The red solid lines are the median relation estimated by the redshift-averaged hierarchical Bayesian models. We also show in the \prospector{} plot the median relationship for the simple model as a reference line. The shaded regions show the 1$\sigma$ and 2$\sigma$ range of the model posteriors. The two models show similar median relationships from the HBM despite the significant differences in their model assumptions. The \prospector{} model has greater scatter around the median relation than the simple model. The simple HBM fits are not able to describe the behavior at the bluest colors.}
    \label{fig:posteriors}
\end{figure*}

Figure~\ref{fig:posteriors} shows the \gr{}--$\logML$ relation for the simple and \prospector{} models. We present the average relationship from the priors, the individual SED fits, and the distribution derived from our HBM. We marginalize the HBM over redshift when we project the 3D model distribution onto the 2D \gr{}--$\logML$ plane. 
Table~\ref{tab:modelparams} lists the posteriors of the 14-parameter Student's t-distribution that is used to describe the distribution of $\logML$ at fixed color. The model parameters are overall well constrained except the quadratic parameters of the kurtosis $q$. 

\begin{figure*}
    \centering
    \includegraphics[width=\textwidth]{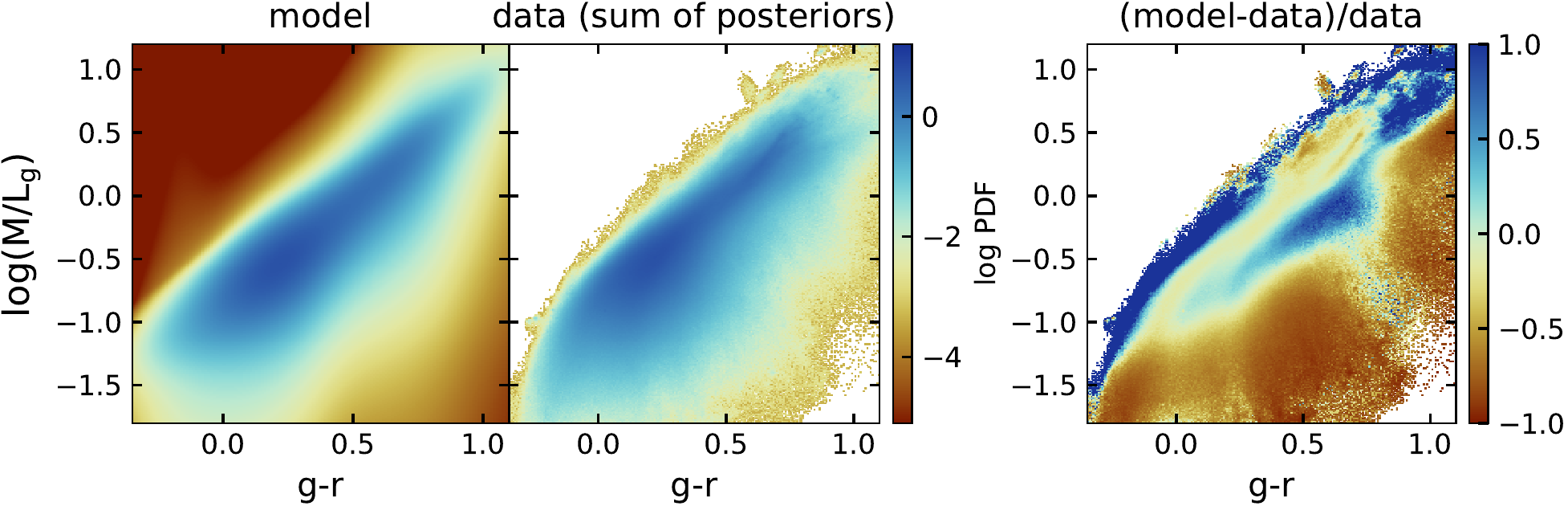}
    \caption{Comparison between the HBM posteriors and the SED-fitting posteriors for the \prospector{} model. Left: the logarithmic density distribution of HBM posteriors marginalized by the redshift and color distribution of the data. It is also convolved with data uncertainties for comparing to the data directly. Middle: the logarithmic density distribution of the sum of \prospector{} SED-fitting posteriors. Right: the fractional distribution difference between the HBM and SED-fitting posteriors. Note that this is not the quantity that our HBM minimizes, and we only use this figure to visualize how well our HBM describes the data. The comparison shows that the HBM performs good fits where the most data are, but is not able to capture the behavior of the upper edge of the \gr{}--$\logML$ distribution where there is little data.}
    \label{fig:residuals}
\end{figure*}

To verify that our HBM fits the data well, we compare the \gr{}--$\logML$ posterior distribution from the HBM and the sum of SED-fitting posteriors of the entire sample for the \prospector{} model in Figure~\ref{fig:residuals}. Because the HBM input is SED-fitting likelihoods for the whole sample, we need to compare the full posterior distribution instead of point estimates, which do not represent the correlated, high-dimensional posterior distribution on redshift, \gr{}, and $\logML$. It is challenging to define a ``residual" between the data, i.e., SED-fit posteriors, and a best-fit HBM model and nor did the HBM aim to minimize any kind of residuals. We processed the HBM posterior distribution in two steps to make it comparable to the data. First, we weighted the HBM posteriors by the redshift and \gr{} distribution of the data to marginalize the 1D $\logML$ distribution at given redshift and color. Second, we convolved the weighted HBM posteriors with the median uncertainties of the observed \gr{} and $\logML$ in each grid of Figure~\ref{fig:residuals} using Gaussian kernels. The final product is shown in the left column of Figure~\ref{fig:residuals}.

In the right column of Figure~\ref{fig:residuals}, we use the fractional difference between our model posteriors described in the last paragraph and the SED-fitting posterior sum to diagnose how well the HBM fits the data. We observe a close match between the model and data distribution density where the most data resides, as shown by the light-yellow color. This result suggests that our HBM works as expected. Since the population model is constrained by the observed galaxies, it should agree with the observed galaxies in the region where most of them reside. The difference is large where there is little data, especially for the upper edge of the relationship where the HBM posterior has a much higher density in the deep blue region than the data posterior sum. The comparison indicates that our HBM describes the main trend of the \gr{}--$\logML$ relationship well but is not able to capture the behavior of some outliers.

In Figure~\ref{fig:posteriors}, we find that the two physical models have similar average relationships from the HBM (see the red and magenta lines in the right column). Though the simple relationship has a larger curvature, in which the slope is steep for blue galaxies and then flattens for red galaxies. In the contrast, \prospector{} has a roughly fixed slope for all galaxies. 
Based on Equation~\ref{eq:Parametrization} and the median posteriors of $a_{0,1,2,3}$ in Table~\ref{tab:modelparams}, the mode of the simple relationship is
\begin{equation}\label{eq:simplerelationship}
    \logML = -0.891 + 2.068(\textsl{g}-r) - 0.503(\textsl{g}-r)^2 - 0.019 z;
\end{equation}
and the mode of the \prospector{} relationship is
\begin{equation}\label{eq:prospectorrelationship}
    \logML = -0.659 + 1.541(\textsl{g}-r) + 0.149(\textsl{g}-r)^2 - 0.121 z.
\end{equation}
Readers may use this equation to estimate the \gr{}--$\logML$ relationship from this work. Note that the \gr{} color here refers to the median from SED-fitting posteriors (see Section~\ref{sec:deltagr} for a discussion on different methods for color measurement). We report the mode of the relationship instead of the mean here to show where most galaxies are. The mean and mode have a typical difference around 0.1\,dex because the $\logML$ distribution is skewed at given \gr{}. We warn the readers that our relationship is untested at $z < 0.5$. Also, in Section~\ref{sec:addri} we provide a tentative way to reduce the scatter of the relationship using an additional \ri{} color (Equation~\ref{eq:ri-residual}). 

At fixed color, both models show an increasingly skewed distribution at lower $\logML$. This trend is strongest for blue galaxies with a $\lambda$ parameter closer to -1 (see Table~\ref{tab:modelparams}). Because the distribution is skewed, it is more challenging to estimate the \stellarML{} of blue galaxies, and can result in a high bias if the skew is unknown. 

The most notable difference between the two SED models is the scatter around the average relation. The dispersion of the relationship is significantly larger for the \prospector{} model, which is evident by the larger posterior estimates of the $\sigma$ parameter in Table~\ref{tab:modelparams}. As we introduced in Section \ref{sec:introduction}, the relationship exists in the first place because the effects of dust, age, and metallicity canceled out: variations in these parameters move the galaxies more or less parallel to the main relationship. Nevertheless, our results suggest that the net effect of these variations can alter the direction of the change in the \gr{}--$\logML$ plane when we consider a more realistic model. 
Hence we need to be careful about using a simple linear relation to predict $\log M/L$ ratios from optical colors.

A distinctive feature of the simple SED fits is the hook at the bluest end. This is due to the similar SFHs of the bluest galaxies in the simple fits, marked by the large $t_\mathrm{f}/t_\mathrm{age}$. As found by \citet{Bell&deJong2001}, recent bursts of star formation will bias \stellarML{} to low values. The exponentially declining SFH suffers strongly from the ``outshining" effect \citep{Papovich2001, Maraston2010}, where the young bright stars outshine the old dim ones in the emitted light. As a result, the exponentially decaying SFH only characterizes the bulk of the most recent star formation and is not sensitive to underlying old populations with recent bursts. The simple fits may cause old  star-forming galaxies to look like young star-forming galaxies in this sense. The lack of flexibility of the SFH model creates a bias on the \gr{}--$\logML$ relationship at the bluest colors.

The principal point to be made here is that we cannot constrain the \gr{}--$\logML$ relation with the simple exponentially declining SFHs. The hook at the bluest color also makes the simple fits difficult to model. As shown in Figure~\ref{fig:posteriors}, our median relation from the HBM is higher than the SED fits for these blue galaxies since the HBM causes regression toward the mean. The diagonal edge of the simple fits is due to the lower bound of $t_\mathrm{age}$ prior, in which the youngest age allowed in the simple model is 10\,Myr. 


\begin{deluxetable}{ccc}
    \tablenum{1}
    \tablecaption{1-$\sigma$ Posteriors for the hierarchical model parameters}\label{tab:modelparams}
    \tablewidth{0pt}
    \tablehead{
    \colhead{Parameter} & \colhead{Simple} & \colhead{\prospector{}} 
    }
    \startdata
    $a_0$ & $-0.8912^{+0.0004}_{-0.0004}$ & $-0.659^{+0.002}_{-0.002}$\\
    $a_1$ & $2.068^{+0.004}_{-0.003}$ & $1.541^{+0.008}_{-0.006}$\\
    $a_2$ & $-0.503^{+0.004}_{-0.005}$ & $0.149^{+0.008}_{-0.010}$\\
    $a_3$ & $-0.0190^{+0.0002}_{-0.0002}$ & $-0.121^{+0.001}_{-0.001}$\\
    $\lambda_{-0.3}$ & $-0.999^{+0.002}_{-0.001}$ & $-0.989^{+0.006}_{-0.006}$\\
    $\lambda_{0.5}$ & $-0.450^{+0.006}_{-0.006}$ & $-0.451^{+0.005}_{-0.005}$\\
    $\lambda_{1.3}$ & $0.995^{+0.003}_{-0.006}$ & $-0.330^{+0.027}_{-0.020}$\\
    $p_{-0.3}$ & $0.413^{+0.005}_{-0.005}$ & $1.259^{+0.063}_{-0.123}$\\
    $p_{0.5}$ & $0.604^{+0.012}_{-0.015}$ & $0.826^{+0.022}_{-0.036}$\\
    $p_{1.3}$ & $0.372^{+0.010}_{-0.011}$ & $0.512^{+0.007}_{-0.007}$\\
    $q_{-0.3}$ & $79.267^{+13.187}_{-5.987}$ & $0.360^{+0.088}_{-0.088}$\\
    $q_{0.5}$ & $5.508^{+0.900}_{-0.544}$ & $19.653^{+5.260}_{-4.332}$\\
    $q_{1.3}$ & $139.118^{+27.817}_{-14.611}$ & $77.119^{+23.018}_{-19.031}$\\
    $\sigma$ & $0.0100^{+0.0005}_{-0.0006}$ & $0.062^{+0.003}_{-0.004}$\\
    \enddata
\end{deluxetable}

\subsection{Redshift Evolution of the Color-$M/L$ Relationships}
\begin{figure}
    \centering
    \includegraphics[width=\columnwidth]{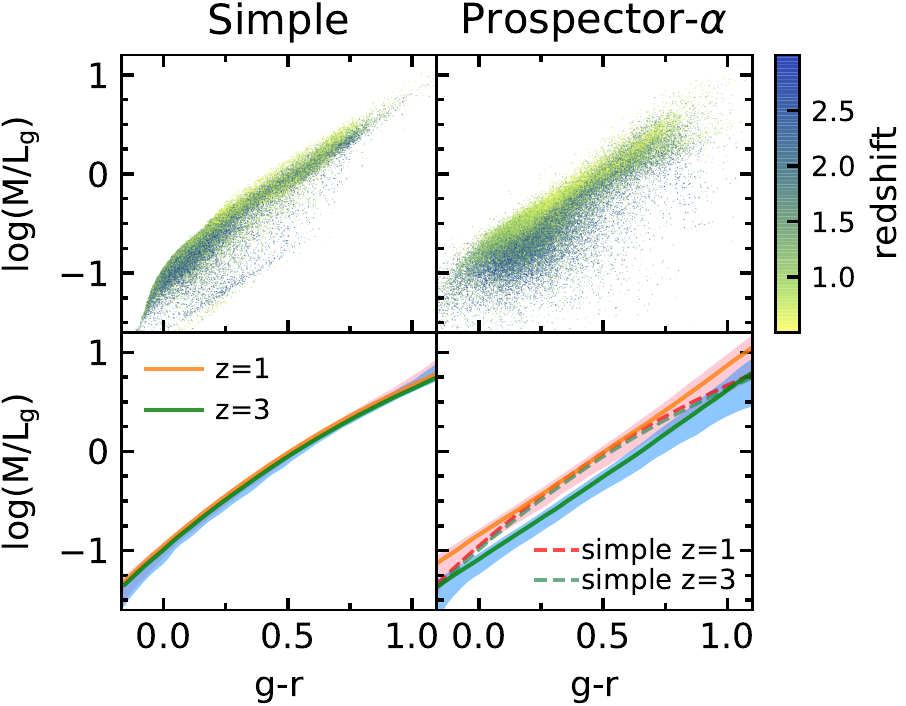}
    \caption{The redshift dependence of the color--\stellarML{} relationship.
    The \prospector{} relationship evolves with redshift while the simple relationship does not. The top panels show the data from SED fits color-coded by redshift. The bottom panels show the HBM results. The solid lines represent the median relations from the HBMs. The shaded regions cover the 16th and 84th percentiles of the posterior distribution. The dashed lines in the lower right plot show the simple relationships, as a comparison to the \prospector{} relationships represented by solid lines.}
    \label{fig:zevolv}
\end{figure}

In addition to the color dependence discussed above, our HBM parameterizes the redshift evolution, prescribed by the $a_3$ parameter (see Table~\ref{tab:modelparams}). The bottom panel of Figure~\ref{fig:zevolv} demonstrates the redshift evolution of the model relationships from $z=1$ to $z=3$. 
The simple model has no perceptible redshift evolution. Conversely, as the redshift increases, the \prospector{} relationship shifts downwards. At given \gr{}, low-$z$ galaxies have higher \stellarML{} than those of high-$z$ galaxies because the stars in these galaxies are on average older. The slope of \prospector{} relationship does not vary as a function of redshift, which means the redshift evolution produces a similar effect for galaxies at all colors.

The strong redshift dependence of the \prospector{} \gr{}--$\logML$ relationship matches simple expectation from passive stellar aging: it increases the average age of the stars, and thus their $M_*/L$ as the age of the universe increases. The simple relationship agrees with the \prospector{} relationship at the blue and red ends at $z=3$, but agrees with the main locus of the \prospector{} relationship at $z=1$. The weak redshift dependence of the simple model implies that it does not reproduce a simple expectation that stars in galaxies get older as the universe gets older.  
This is because exponentially declining SFH only models the light that it sees from bright stars and is not sensitive to old stellar populations as we discussed in Section~\ref{sec:HBMrelation}. 
Previous studies such as \citet{ALHAMBRA2019} found no change in the relationship with $z$, which is also likely due to their use of parametric SFHs.


\subsection{What Drives the \stellarML{} Offset Between the Two Models} \label{sec:deltaML}
So far we have shown how the optical color--\stellarML{} relationship derived from two SED models differs. According to \citet{prospector-massfunc}, \prospector{} provides one of the highest \stellarML{} from fitting the photometry among various SED models, which is still true when fitting the spectrum \citep{Tacchella2022}. 
The next question is what physics properties drive the differences between the models. Much of the complexity in interpreting the relationship is that the model parameters correlate with each other. To disentangle the degeneracy quantitatively and identify the most relevant driver of the higher inferred \stellarML{} values, we exploit the feature importance measurement provided by random forests (RFs).

We train an RF to predict the $\logML$ difference between the \prospector{} model and the simple model. We use the {\sl scikit-learn} package \citep{Pedregosa2011scikit} to construct an RF model of 1000 trees. We adopt 80\% of our sample as the training set, and 20\% as the test set. The training features are extracted from the SED fits, which include the model differences in the \gr{}, diffuse dust optical depth $\hat\tau_2$, total formed mass-weighted age $t_\textrm{avg}$, and SFR in recent 100\,Myr $\textrm{SFR}_\textrm{100\,Myr}$, as well as the \prospector{} measurements of \gr{}, $\hat\tau_2$, $t_\textrm{avg}$, specific star formation rate $\textrm{sSFR}_\textrm{100\,Myr}$, stellar metallicity, and gas-phase metallicity. The endpoint of the RF regression is an accuracy score to reflect the goodness of fit, with 1 being the most accurate. Our selected features can predict $\Delta \logML$ of the test set at a mean accuracy score of 91\% with a 0.03\,dex root-mean-square error. The prediction accuracy is even better if we only consider high signal-to-noise ratio ($S/N$) galaxies. This suggests that we have included sufficient training features. 

\begin{figure}
    \centering
    \includegraphics[width=\columnwidth]{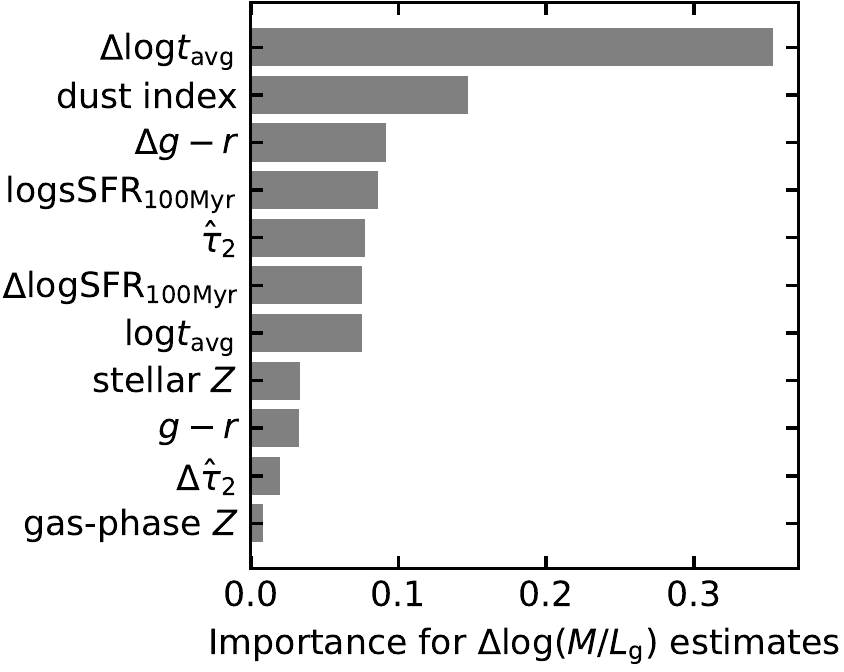}
    \caption{The importance of different factors in predicting the $\logML$ difference between the simple and \prospector{} model. The age difference is the most important.}
    \label{fig:rf}
\end{figure}

Figure~\ref{fig:rf} shows the importance of each input feature, which is computed by the reduction of the model performance by dropping out the features. Figure~\ref{fig:rf} immediately demonstrates that the difference in mass-weighted age is the primary driven factor for the $\logML$ offsets. $t_\textrm{avg}$ is far more predictive than the other features with a relative importance of 35\%. This means that \prospector{} fits have systematically higher $\ML$ ratios mostly because they are older, as a result of using nonparametric SFHs (see \citealt{Lower2020} for a discussion on the $M_*$ estimates from different SFHs). This agrees with the redshift trend in Figure~\ref{fig:zevolv}. The bulk of the difference between the model relationships comes from galaxies at lower redshifts because the age differences permitted are larger \citep{2019ApJ...877..140L}. The principal role of age is also supported by studies based on SPS libraries. For example, \citet{Into2013} demonstrated that both colors and $M/L$ ratios increase continuously with SSP age regardless of the metallicities. 

Other factors exhibit smaller influences than $t_\textrm{avg}$ on $\Delta \logML$ estimates, with the dust index being the second important feature. The dust index is a power-law modification to the shape of \citet{Calzetti2000} dust attenuation curve \citep{Noll2009}, where Calzetti law has a dust index of 0. Galaxies with a larger dust index have a flatter curve, i.e., less attenuation in UV and more attenuation in NIR. The dust index does not directly affect $L_\textsl{g}$ because $\hat\tau_2$ is measured within the $\textsl{g}$ band, but will influence $M_*$, and thus $M_*/L_\textsl{g}$. As the NIR SED traces closely the bulk of $M_*$ from the old, low-mass stars, a large dust index permits galaxies to add a lot of mass due to increased attenuation in NIR without changing optical colors \citep{Salmon2016, Malek2018}. 

We find that the dust index is the most distinguishing factor for red galaxies when we repeat the RF analysis for the red subsample.
Since dusty galaxies on average have a flatter dust attenuation curve than galaxies with little dust \citep{Chevallard2013, dustlaw2020review}, having a free dust index allows red galaxies to be either dusty star-forming galaxies with flatter dust attenuation curves (e.g. ultraluminous infrared galaxies), or nearly dust-free quiescent galaxies with steeper curves. In contrast, the simple model assumes a zero dust index for all galaxies. The distinct dust indices between the two models lead to large $\logML$ offsets at the red end.

The sSFR also has a second-order effect on $\Delta \logML$. The simple fits do not model the dust emission in IR while the IR emission traces the recent star formation. This generally biases the SFRs to lower values since dust-obscured star formation is preferentially missed \citep{Wuyts2011}. Our results confirm that the $\logML$ offsets are marginally influenced by metallicity, especially gas-phase metallicity.


\subsection{Color Systematics}\label{sec:deltagr}

The results discussed so far investigate one component of the relationship, the mass-to-light ratio. 
The other part, the rest-frame optical color, is usually considered a well-measured quantity. Nevertheless, systematics occurs when we convert the observed photometry at different redshifts to the rest-frame colors. In our work we synthesize colors from the models generated during SED fitting.
Some studies and surveys adopt K-corrections to convert the observed-frame photometry of one given band to the desired rest-frame photometry of the same or a different band (e.g, \citealt{Hogg2002}; \citealt{Blanton2007}). K-correction from the observed band to its rest-frame counterpart is an efficient way to estimate colors at low redshift presumably since the bandpass wavelengths have changed very little. As one goes to high redshift, the dependence of the same band K-corrections on the galaxy SED increases but some surveys try to use an observed bandpass near the blueshifted rest-frame bandpass to approximate. K-corrections are usually calculated by fitting the observed photometry with linear combinations of a limited number of SED templates (\citealt{Blanton2007}; some also use a single best-fit SSP to calculate K-corrections \citep{Chilingarian2010}). Although the calculation of K-corrections still involves SED fitting at some level, the resultant color is likely to be less model-dependent than the one from full SED fitting. But unlike K-corrections, colors from SED fitting can be easily calculated in a consistent fashion for galaxies at any redshifts. Also, SED fitting can infer other galaxy properties including ages, metallicities, dust properties, and detailed SFHs more than just colors.

\begin{figure*}
    \centering
    \includegraphics[width=\textwidth]{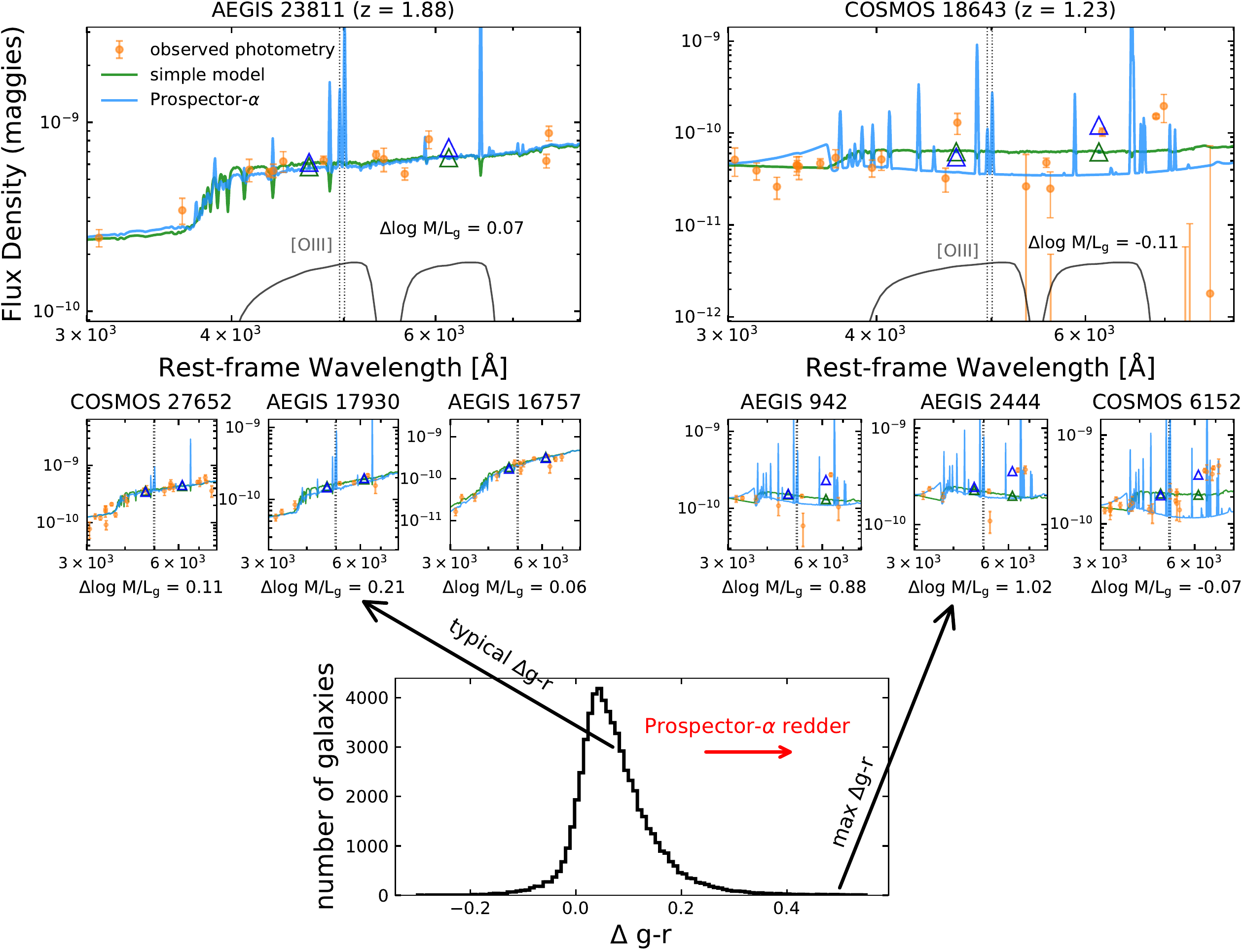}
    \caption{The color differences between the two models. The bottom panel shows the distribution of the \gr{} offsets between \prospector{} fits and the simple fits, where the majority of galaxies are redder in \texttt {Prospector-$\alpha$}. In the top panels, we show example spectra of galaxies with the typical color difference and the extreme color difference, where $\textsl{g}$-band and $\textsl{r}$-band model photometries are shown in triangles. Comparison between the two columns in the top panels indicates that the model offset in the rest-frame color mainly comes from the nebula emission.}
    \label{fig:colordiff}
\end{figure*}

In Figure~\ref{fig:colordiff}, we contrast the median model rest-frame colors and spectra of several example galaxies between the two SED model fits. The typical color differences between the models are around 0.1\,mag, with the \prospector{} results systematically redder. In the upper left panel, we show galaxies with the typical color differences, where the corresponding $\textsl{g}$-band and $\textsl{r}$-band fluxes are shown as triangles. In the upper right panel, we show examples of galaxies with the most extreme color differences (all $>0.5$\,mag). 

The comparison makes it clear that the color offsets between the two models are mostly due to the emission lines. The simple model does not include nebular emission. In particular, if the galaxy has bright UV fluxes and shows no prominent Balmer break, the simple model cannot fit the data well because the Balmer break is its only sharp spectral feature. Conversely, \prospector{} is more flexible in modeling observed photometry with large variations. It will add strong emission lines in this situation, which produces a lower $\chi^2$ value when fitting the photometry as compared to the simple model (see upper right panel of Figure~\ref{fig:colordiff}). Combining our discussion in the last section, the shift of $\logML$ with a median 0.12\,dex due to nonparametric SFHs and the shift of \gr{} with a median 0.06\,mag due to nebular emission, together move the average \prospector{} relationship in a parallel direction as the average simple relationship.  

Remarkably, the galaxies with the extreme color differences in Figure~\ref{fig:colordiff} all show the inverse Balmer break feature. The inverse Balmer break comes from strong nebular continuum emission. This feature is uncommon for galaxies at lower redshifts but more likely to be prevalent in the early universe, marked by very high SFRs, strong nebular emissions, and very low metallicities. Our finding suggests that a subsample of \hst{} galaxies are candidates for being extreme emission line galaxies (EELGs, e.g., \citealt{Maseda2018}). We examine the H$\alpha$ equivalent width (EW) for 20 galaxies with the most extreme color differences. All of them have H$\alpha$ EW $\gtrsim 400$\,km $\mathrm{s}^{-1}$, high sSFR, and low metallicity, consistent with the properties of observed EELGs.

\subsection{Shrinkage Effects in the HB method}\label{sec:HBMshrinkage}
Here we will discuss a key feature of HBMs called shrinkage -- this refers to the reduction in posterior size that occurs when applying the population model as a prior to individual fits. 
In our problem, the shrinkage estimators introduce a decrease in variance around the average \gr{}--$\logML$ relationship. Because of the restrictions of the simple model and its curvature at the bluest color that makes it hard to model, we will focus on the \prospector{} model for the shrinkage analysis in this section. We calculate the shrinkage of every galaxy by multiplying the individual SED-fit posterior $P(\btheta_i|\bD)$ by a factor $P(\brho_i|\bD)/P(\btheta_i)$, where $P(\brho_i|\bD)$ is the HBM posterior at this galaxy's redshift and $P(\btheta_i)$ is the prior probability we show in Figure~\ref{fig:priors}. In this way, we are reapplying the HBM posterior as a prior to the individual likelihoods from SED fitting.

\begin{figure*}
    \centering
    \includegraphics[width=.85\textwidth]{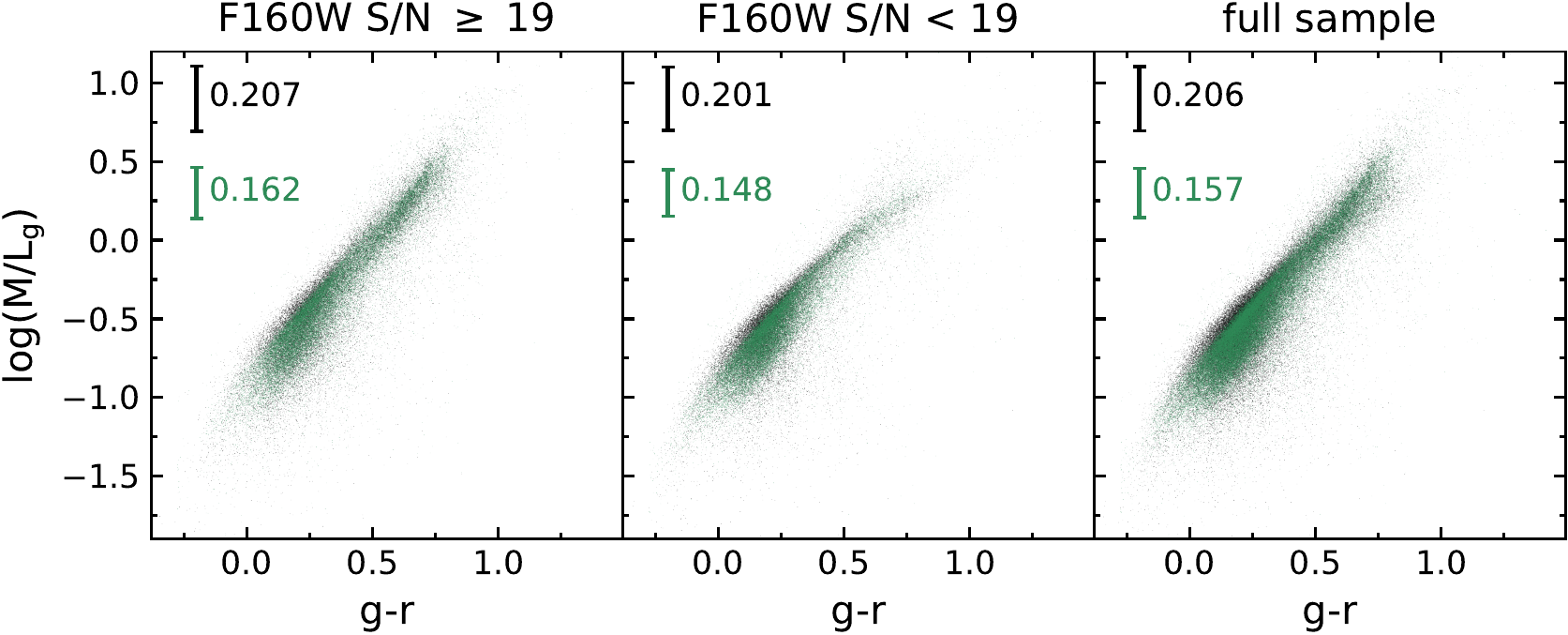}
    \caption{Comparison of the \prospector{} fits to the HBM prediction after shrinkage for $S/N \geq 19$ galaxies, $S/N < 19$ galaxies, and the entire sample. Galaxies before the shrinkage are indicated in black, and galaxies after shrinkage are shown in green. The error bars represent the median standard deviation of $\logML$ over all colors.
    The model does stronger shrinkage for low $S/N$ galaxies.}
    \label{fig:shrinkage}
\end{figure*}

Figure~\ref{fig:shrinkage} presents the shrinkage effects on the \prospector{} model. We compare the posterior mean of each galaxy before and after shrinkage, where the original SED fits are plotted in black and the data after shrinkage is in green. A close inspection of this shows a more concentrated posterior distribution after shrinkage. The model shifts most galaxies with high $\logML$ toward the median relationship. We compare the shrinkage effects between galaxies with different $S/N$ ratios in different panels. The full sample is divided into two subsamples using $S/N = 19$ as a cut, by which we have a roughly equal number of galaxies in each subsample. It is clear that the individual fits experience stronger shrinkage in the $S/N < 19$ subsample, where the galaxies are dimmer. Because such faint galaxies have a smaller effect on the inferred hyperparameters and we assume the faint galaxies distribute similarly to the bright galaxies, we are borrowing information from bright galaxies when we perform hierarchical shrinkage on the properties of faint galaxies. This manifests a strength of the HBM that it provides a framework whereby a subset with high-quality data explicitly benefits all of the data.

\section{Discussion}
\label{sec:discussion}
Here we have learned that the chosen SED model has a large effect on the optical color, $M_*/L$, and the relationship between the two. Now we will discuss three outstanding issues. First, we will make a comparison to the color--\stellarML{} relationships in other work. Second, we will discuss the implications of the HBM shrinkage. Finally, we will explore the possibility of constraining a much tighter relationship with an additional color \ri{}.

\subsection{Comparison to Other Color--\stellarML{} Relationships}
\begin{figure}
    \centering
    \includegraphics[width=.85\columnwidth]{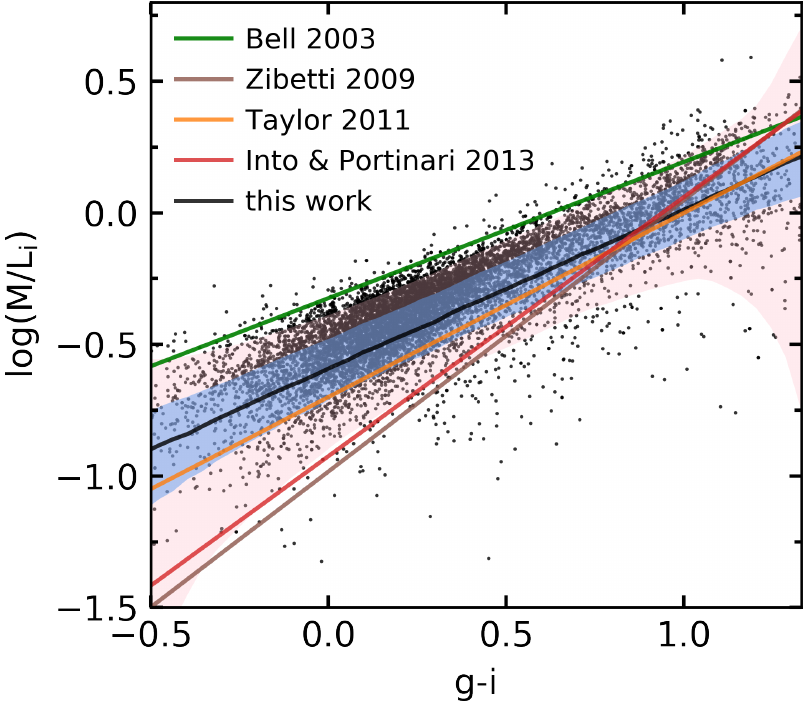}
    \caption{Comparison to the $(\textsl{g}-\textsl{i})$--$\log M/L_\textsl{i}$ relationships presented by other works. The black solid line is \prospector{} median relationship from our HBM. The blue and pink shaded regions are the 1$\sigma$ and 2$\sigma$ range of the model posteriors respectively. The green line shows the relationship from \citet{Bell2003}. The yellow line represents the relationship given by \citet{Taylor2011GAMA}. These two are empirical relationships derived from SED fitting to data like our work. The \citet{Zibetti2009}, \citet{Into2013} relationships are based on stellar evolution models.}
    \label{fig:comparison}
\end{figure}

\prospector{} model predicts higher galaxy masses and smaller SFRs than other SED models \citep{prospector-massfunc} and we expect it to also produce a different color--\stellarML{} relationship. In Figure~\ref{fig:comparison}, we compare the median \prospector{} $(\textsl{g}-\textsl{i})$--$\log M/L_\textsl{i}$ relationship to two empirical relationships and two theoretical relationships in other works. We switch to $(\textsl{g}-\textsl{i})$ and $\log M/L_\textsl{i}$ for consistency with these works. We follow the same procedure in Section \ref{sec:model} to calculate the distribution of $\log M/L_\textsl{i}$ at given $(\textsl{g}-\textsl{i})$ by fitting the HBM. We have accounted for the offsets due to the choice of different IMFs. 

The relationships from \citet{Zibetti2009}, \citet{Into2013} are derived from SPS libraries. \citet{Zibetti2009} estimate their relationship by marginalizing over a Monte Carlo library of SPS models, with an SFH composed of an exponentially declining continuum and random SF bursts. \citet{Into2013} use the Padova isochrones to show the trends in $M/L$ ratios and colors as a function of ages, metallicities, SFH birthrate parameters. These relationships depend entirely on the physical model and should in fact be compared to our priors. Similar to the priors in Figure~\ref{fig:colorpriors}, these relationships have steeper slopes than our relationship from observational data.

The empirical relationships from \citet{Bell2003}, \citet{Taylor2011GAMA} are more aligned with our relationship. \citet{Bell2003} fit the relationship using data from the Sloan Digital Sky Survey (SDSS) and the Two Micron All Sky Survey (2MASS). They use an exponentially declining SFH and evolve their dust-free relationship back to $z=0$ where they assume the galaxies are 12\,Gyr old. Our offset to their relationship can be partly explained by the inclusion of dust, and by the difference in age due to our nonparametric SFH and higher central redshift (see redshift evolution in Figure~\ref{fig:zevolv}). 
\citet{Taylor2011GAMA} obtained the relation using intermediate-redshift ($z<0.65$) galaxies in the Galaxy And Mass Assembly (GAMA) survey with an exponentially declining SFH as well. Their relationship is the closest to ours. The \citet{BC2003} SPS models that they choose can introduce a $\sim$0.05\,dex offset in $M_*$ with respect to FSPS SPS models used in \prospector{} \citep{prospector-massfunc}.

The relationships from literature in Figure~\ref{fig:comparison} are all based on parametric SFHs. We have argued that nonparametric SFHs are crucial to providing the flexibility needed to model \stellarML{} estimates for blue galaxies. Recently, \citet{Ge2021} also use nonparametric SFHs for SED fitting to study the pixel-by-pixel-based color-$M/L$ relationship of MaNGA galaxies. However, their $M/L$ estimates are systematically higher than ours, and we speculate that it is because individual pixels are allowed to have a more bursty SFH. While galaxies represent the integrated light from pixels, the bursty features of pixels are possibly smoothed out. Also note that we assume a continuity SFH prior from \citet{prospector-massfunc} and they assume a much bustier prior from \texttt{PPXF} \citep{PPXF2017}. Therefore, our relationship cannot be compared to theirs directly; this highlights the great effects of the nonparametric SFH priors on the inferred SFH. 

To summarize, there are many ways that \prospector{} color--\stellarML{} relationship may be different than the other empirical results. First, \prospector{} is a sophisticated physical model that adds nebular emission, dust emission, and uses nonparametric SFH. All these model assumptions may affect the relationship. As we have highlighted in Section~\ref{sec:result}, the nonparametric SFH shifts the whole relationship as the stellar population ages. This means if our sample centers at a different redshift than other works do, the resultant average relationship will be different. Second, each work assumes different model priors. Consequently, the parameter space covered in our (other) model may not be allowed in other (our) models. For example, \citet{BC2003} SPS model covers a wider variety of metallicity than ours, which can lead to smaller \stellarML{} for blue galaxies because they are allowed to have lower metallicity. Despite the discrimination between our and other color--\stellarML{} relationships, we do not know which relationship is correct.

Our updated relationship between color and \stellarML{} can potentially impose constraints on galaxy dynamical masses ($\Mdyn$). The difference between dynamical and stellar masses often depends on how far from the galaxy center we probe.  If we have spatially resolved data to take the galaxy radius into account explicitly, we may correlate the $\Mdyn/L$ with the \stellarML{} of a galaxy (e.g., \citealt{Taylor2010}; \citealt{deGraaff2021}). 
For example, \citet{Taylor2010} showed that the ratio of $M_*$ and $\Mdyn$ is independent of stellar population parameters and observables such as apparent magnitude and redshift after excluding the effect of galaxy structure. A few studies observe a strong empirical relationship exists between $\log \Mdyn/L$ and color (e.g., \citealt{van-de-Sande2015}) similar to $\log M_*/L$. Since $\Mdyn$ is composed of stellar, gas, and dark matter components, our relationship on \stellarML{} can set a lower limit of $\Mdyn/L$. Specifically, the relationship can potentially put radius-dependent constraints on $\Mdyn/L$ at high redshift where the galaxy kinematic studies are challenging. Note that there is much complexity in the interpretation of the dynamical masses such as the galaxy inclination and mass--anisotropy degeneracy, and therefore, one needs to be cautious about linking $\Mdyn$ with $M_*$.

\subsection{Learning New SED-fitting Priors from the HBM Shrinkage}
\begin{figure*}
    \centering
    \includegraphics[width=\textwidth]{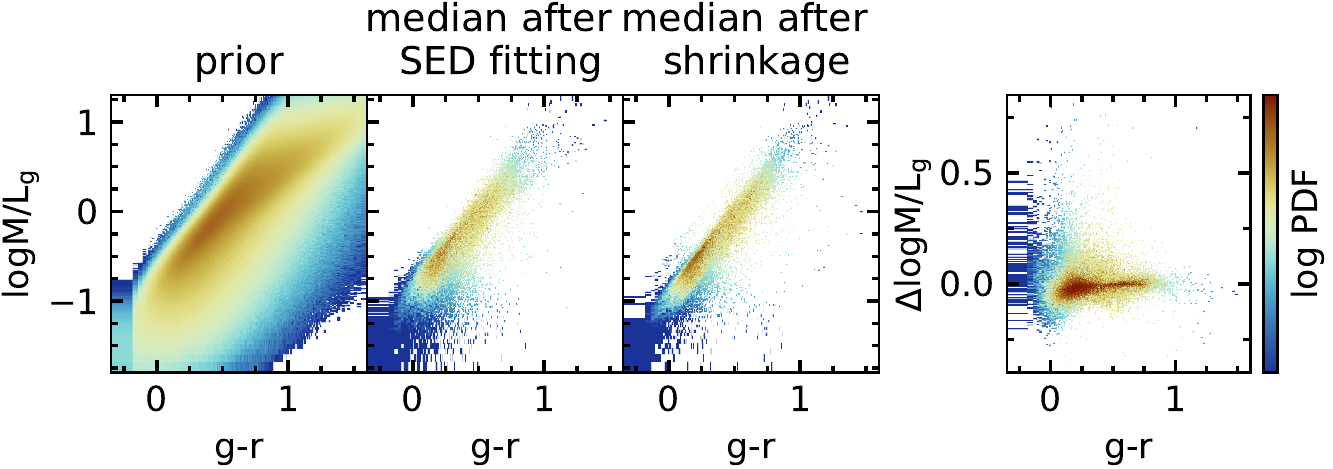}
    \caption{The logarithmic density of galaxies on the \gr{}--$\logML$ plane for the priors, the median of individual SED-fitting posteriors, and the median of each object after HBM shrinkage are shown in the left three columns respectively. The right column demonstrates the $\logML$ offset after and before the HBM shrinkage. The results shown here are based on the \prospector{} model. We observe a strong shrinkage effect for blue galaxies, but some red faint galaxies are pushed away from the average relationship.}
    \label{fig:prior_shrinkage_comparison}
\end{figure*}

While \prospector{} likely allows more freedom in its parameters than galaxies actually occupy, the HBM provides us with one way to update the SED model to make it better describe the real data. A key aspect of the population model is to link galaxies of the same population and facilitate predictions of unobserved galaxies of this population. Potentially, we can infer from the shrinkage estimates how we should tune the priors to make them better fit the underlying population distribution. Deriving new priors from the shrinkage may not be a straightforward process \citep{Loredo&Hendry2019}. But understanding the HBM shrinkage would be a first step to building better priors.

Figure~\ref{fig:prior_shrinkage_comparison} shows the probability density function of priors, the median posteriors from individual SED fits, the median estimates of every galaxy after HBM shrinkage in the \gr{}--$\logML$ plane, and the difference in $\logML$ between the median after shrinkage and the median after SED fitting for each galaxy. For the majority of galaxies, the shrinkage estimates are more tightly constrained to the average relationship than the individual SED fits. A large amount of shrinkage is observed for the blue galaxies where the HBM increases their extremely low $\logML$ values. The shrinkage analysis suggests that most galaxies do live within a narrower range than our priors, especially blue galaxies. If we apply this tighter color--$M/L$ relationship from the HBM for future photometric surveys like the Large Synoptic Survey Telescope (LSST), it may enable a more accurate selection of color and redshift for the mass range covered than those from the current model priors (see Figure~\ref{fig:priors}).

Despite the significant shrinkage at the blue end and in the middle color range shown in Figure~\ref{fig:prior_shrinkage_comparison}, we observe an opposite effect for some red faint galaxies.
For these outliers, the HBM pushes them even farther from the average by lowering their $\logML$. This is because the prior is too strong and washes out the effect of HBM posterior. The different results for blue and red galaxies from the shrinkage analysis may imply that our assumption that faint and bright galaxies are drawn from the same population may not be true, and we need more population components in our model. 
In order to test the efficacy of our model on specific galaxy subpopulations, it will be important to have deeper, higher spectral resolution data across the entire color range or a larger sample size.

The shrinkage analysis also supports our choice of the population model for the distribution of galaxies in the optical color--$\log M_*/L$ plane. We choose a flexible functional form to model the population distribution as described in Section \ref{sec:model}. The purpose of using a flexible model is to avoid biases led by model choices. Adding free parameters into the population model will usually lead to greater shrinkage toward the average relationship as the population model tries to avoid overfitting. Figure~\ref{fig:prior_shrinkage_comparison} only shows a moderate amount of reduction in scatter. Given that the amount of shrinkage is determined by $P(\brho_i|\bD)/P(\btheta_i)$, if galaxies in fact live far from the SED model predictions or the population model is unnecessarily complex, the amount of shrinkage should be much greater.

In addition to learning from data using the hierarchical model, numerical simulation is another approach to building more informed priors \citep{Pacifici2013}. However, galaxy evolution is very complicated, and simulations need to consider many ingredients, such as the interstellar medium, dark matter halo, active galactic nuclei, stellar feedback, gas cooling, and magnetic fields. It is a challenge to ensure that all the physical parameters we are interested in are well predicted in simulations \citep{Somerville&Dave2015}, especially when considering different subgrid physics (e.g., \citealt{Crain2015}).
As a result, each galaxy formation model encodes a different set of SFHs. What we learn from the simulations will vary widely depending on which simulation we choose. For example, the predictions for the power spectrum density have a large diversity among simulations \citep{Iyer2020}, and $M_*$ from SED fitting shows systematic offsets to the true mass from simulated stellar particles \citep{Lower2020}.
Unlike simulations, for HBM we are guaranteed to learn how we can update our SED model to describe the data better but suffer from the observation uncertainties and degeneracies. We can utilize the observations and simulations together to better constrain the SED priors.

\subsection{Constraining the \stellarML{} with Two Colors}\label{sec:addri}

\begin{figure*}
    \centering
    \includegraphics[width=0.9\textwidth]{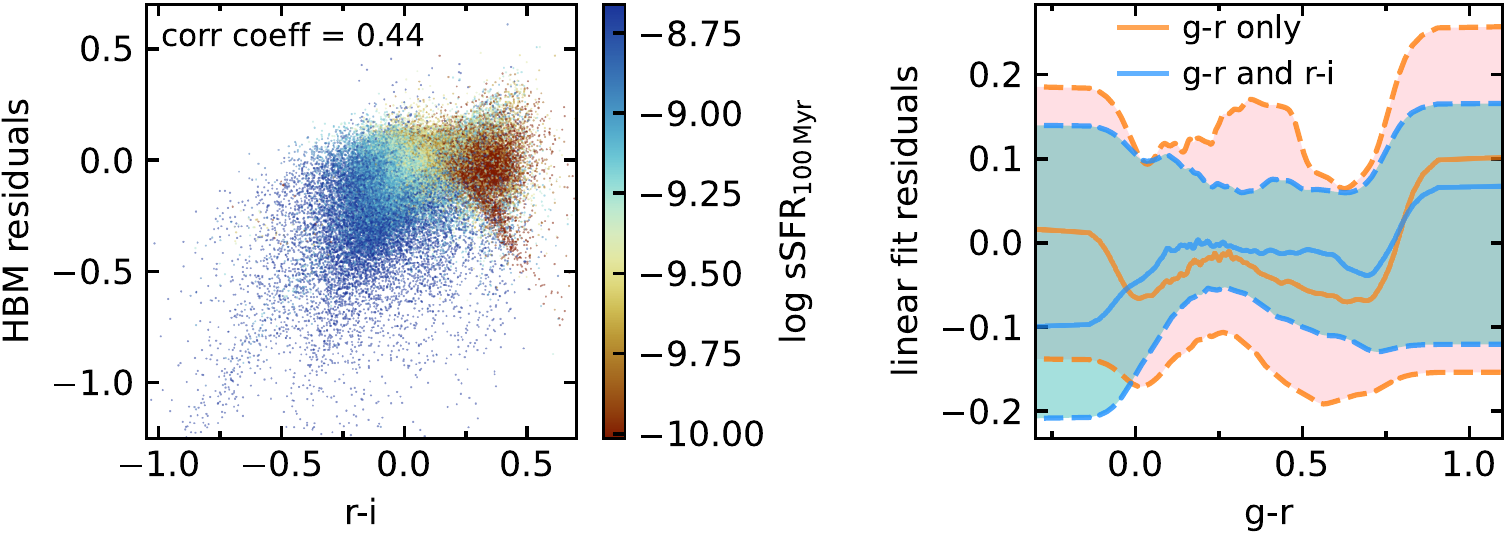}
    \caption{The rest-frame \ri{} color may significantly reduce the uncertainty of \stellarML{} estimates from CMLR. Left: sSFR drives the correlation between \ri{} and $\logML$ residuals calculated from the mode of the HBM relationship. The Pearson correlation coefficient is 0.44. Right: the linear fit of $\logML$ upon both \gr{} and \ri{} has smaller scatter than the linear model using only \gr{}. The shaded regions cover 1$\sigma$ range of the residuals for the two linear models respectively.
    Both plots are based on the \prospector{} model.}
    \label{fig:ri}
\end{figure*}

So far our discussion has been focused on inferring \stellarML{} from one optical color, but is it possible to significantly reduce the uncertainty of \stellarML{} estimates with one or two more colors? In principle, the derived \stellarML{} will certainly be closer to the SED-fit results when we use more colors, or equivalently, more photometry as this is what the SED models are conditioned upon. Our question is if there is one color that provides significant added information on the $\ML$ in addition \gr{}, so the combination of the two colors can constrain a much tighter relationship with $\ML$.

We calculate rest-frame colors from 10 filters in the optical and NIR, including \textsl{u, g, r, i, z, J, H, K, IRAC1, IRAC2}. We examine if any of them are correlated with the $\logML$ residuals between individual SED fits and the HBM \gr{}--$\logML$ relationship from Equation~\ref{eq:prospectorrelationship}. Of these colors, we find that \ri{} has the strongest correlation with the residuals (see the left column of Figure~\ref{fig:ri}). The Pearson correlation coefficient is 0.44. This correlation indicates that including \ri{} may help reduce the residuals of $\logML$ estimates. We perform a linear regression between \ri{} and the residuals of $\logML$ estimates. Readers may use the following relation to evaluate an \ri{} dependent correction to the $\logML$ estimates from our HBM. 
\begin{equation}\label{eq:ri-residual}
    \logML~\textrm{residuals} = -0.11 + 0.40 (\textsl{r}-\textit{i}).
\end{equation} 
Note that this relation has not been tested through our full HBM machinery.

As a simple test to quantify the improvement of \stellarML{} estimates after adding \ri{}, we perform two linear fits on the relationship between color(s) and $\logML$ using the median posteriors from SED fits. In one of the linear models, $\logML$ has a linear dependence on \gr{} and redshift, and in the other model, $\logML$ depends on \gr{}, \ri{}, and redshift. 
The two models predict a similar average relationship between \gr{} and $\logML$ while the model with \ri{} has a smaller scatter around the regression line. The right column of Figure~\ref{fig:ri} shows the median and 1$\sigma$ range of the $\logML$ residuals of the linear fits. The linear model with two colors has residuals closer to 0 and a smaller scatter $\sim$0.15\,dex, compared to $\sim$0.26\,dex of the model with only \gr{}.
Our results imply that it may be worthwhile taking \ri{} into account besides a single color \gr{} to better constrain \stellarML{}. We do not perform a full HBM here since the purpose of this section is only to provide a possible path forward for future analyses.

\ri{} is potentially useful for constraining a tighter color--\stellarML{} relationship because it helps to identify sSFR, which can be degenerate with metallicity or dust at a fixed \gr{}. In the left column of Figure~\ref{fig:ri}, we show that both $\logML$ residuals and \ri{} correlates with $\textrm{sSFR}_\textrm{100\,Myr}$, which means that sSFR is the major driver for the potential of \ri{} in reducing \stellarML{} residuals. At a fixed \gr{} color, a bluer \ri{} color is associated with younger star-forming galaxies, whereas a redder \ri{} color is typically associated with older quiescent systems. We speculate that \ri{} correlates with $\textrm{sSFR}_\textrm{100\,Myr}$ because the \textsl{r}-band photometry likely puts constraints on the H$\rm{\alpha}$ emission line, an indicator of the most recent SFR. If this is true, adding \ri{} may only benefit the color--\stellarML{} relationship from SED models including nebular emission.

The right column of Figure~\ref{fig:ri} indicates that including \ri{} has the largest effect on red galaxies with $(\textsl{g}-\textsl{r}) \gtrsim 0.3$. So \ri{} may be useful in differentiating the dusty star-forming galaxies with blue \ri{} and the quiescent galaxies with red \ri{} for galaxies with red \gr{} color. 
This agrees with our finding that \ri{} has a strong correlation with $\hat\tau_2$.
The fact that \ri{} anticorrelates with sSFR and reduces the scatter of $\logML$ estimates for red galaxies likely suggests that \ri{} helps distinguish between age and dust. 

\section{Summary}
\label{sec:summary}
The empirical rest-frame optical color--\stellarML{} relationship is greatly influenced by the physical model used in SED fits. In this work, we contrast the \gr{}--$\logML$ relationship between a simple 4-parameter SED model and a sophisticated 14-parameter \prospector{} model using 63,430 \hst{} galaxies up to $z=3$. We utilize a hierarchical Bayesian model to fit the relationship, which allows us to account for the SED-fit posteriors and priors explicitly. We show the distinction between the two model relationships and identify the galaxy properties that drive the difference. Furthermore, by taking advantage of HBM shrinkage, we learn more about the true distribution of the \gr{} color and $\ML$.
Our main findings are as follows:
\begin{itemize}
    \item The two contrasting SED models derive similar average \gr{}--$\logML$ relationships but with significantly different scatters. 
    The more sophisticated \prospector{} model predicts a looser and less skewed relationship in contrast to the simple model (Figure~\ref{fig:posteriors}), with a 1$\sigma$ uncertainty of 0.28\,dex compared to 0.12\,dex. 
    \stellarML{} and \gr{} offsets between the two SED models are mostly attributed to the nonparametric SFHs and nebular emission in \texttt {Prospector-$\alpha$}, respectively. However, the two effects together shift the average relationship in a parallel direction to higher \stellarML{} and redder color. The simple model is too restricted and not sufficient for describing all galaxies, especially the blue ones because the exponentially decaying SFH does not allow galaxies to have both old stars and optically blue colors.  
    
    \item Stellar age is the major driver of the difference between two model relationships (Figure~\ref{fig:rf}). For the \prospector{} model, we find a significant redshift evolution of the relationship (Figure~\ref{fig:zevolv}). The average relationship for low-$z$, older galaxies has higher \stellarML{} than that of high-$z$, younger galaxies. \prospector{} nonparametric SFH adds more scatter to the relationship through a more flexible treatment of SFHs. We do not recover a redshift evolution in the simple model due to the limitation of exponentially declining SFHs against outshining effects. The shape of the dust attenuation curve is an important driver for the \stellarML{} offset in red galaxies.
    
    \item The nebular line and continuum emission enabled in \prospector{} is the main reason for the different color measurements (Figure~\ref{fig:colordiff}). The objects with extreme color offsets are likely to be EELGs.
    
    \item The shrinkage analysis of the HBM motivates us to learn better galaxy SED-fitting priors from the data. Our current priors are slightly wider for most galaxies but should be weaker for some red galaxies (Figure~\ref{fig:prior_shrinkage_comparison}). 
\end{itemize}

Our work shows that optical color--\stellarML{} relationship suffers from systematic errors depending on the choice of the SED model. The skewness of the distribution of \stellarML{} at given color may cause a bias when estimating $M_*$ for a specific galaxy using the average relationship. Given the redshift evolution we found, future work needs to be careful when applying the empirical relationship determined locally to high-redshift galaxies. 

Our results can be expanded in several ways. The HBM in this work is a path forward to informing better priors for SED fitting using various hierarchical models. Dynamical masses will provide useful constraints to the relationship, and it may help to include dynamical measurements in the HBM. Ideally, if we have a more flexible model such as an HBM fitting both M/L distribution at a given color and color distribution at a given M/L, we can better parameterize the M/L distribution at different colors. 
Finally, it is possible that the scatter in the relationship can be significantly reduced by adding only one or two more colors such as the $r-i$ color.

\acknowledgments

We thank Eric Bell for an insightful review. We thank Marijn Franx, Charlie Conroy, and Ben Johnson for help in the early stages of this project. We further thank Rachel Bezanson for thoughtful comments and discussion on future science. Also thanks to Pieter van Dokkum for helpful discussion. Computations for this research were performed on the Pennsylvania State University's Institute for Computational and Data Sciences' Roar supercomputer.

\software{Prospector \citep{Prospector-1, Prospector-2}, FSPS \citep{Conroy2009},  PYTHON-FSPS \citep{python-fsps}, dynesty \citep{dynesty}, MIST \citep{MIST0, MIST1}, MILES \citep{MILES}, NumPy \citep{harris2020array}, SciPy \citep{2020SciPy-NMeth}, scikit-learn \citep{Pedregosa2011scikit}, Matplotlib \citep{Hunter2007}, Astropy \citep{astropy:2013, astropy:2018}. \\~\\}

\appendix
\section{Mock Test}
\label{sec:appendixA}
We perform mock tests to validate our population model described in Equation~\ref{eq:ProbML}.  We generate $10^5$ mock galaxies and model them with the sgt distribution discussed in Section~\ref{sec:model}.
The \gr{} color is sampled from a truncated normal distribution, which is similar to the data distribution in Figure~\ref{fig:priors}. For each object, we choose a random redshift between 0.5 and 3. We then generate their $\logML$ from the most likely parameters of the HBM modeling of the \prospector{} SED fits. We build 15 samples for each object, and for each sample we add small intrinsic errors drawn from normal distributions with a scale of 0.01 mag for \gr{} and 0.01\,dex for $\logML$. 

\begin{figure*}
    \centering
    \includegraphics[width=\textwidth]{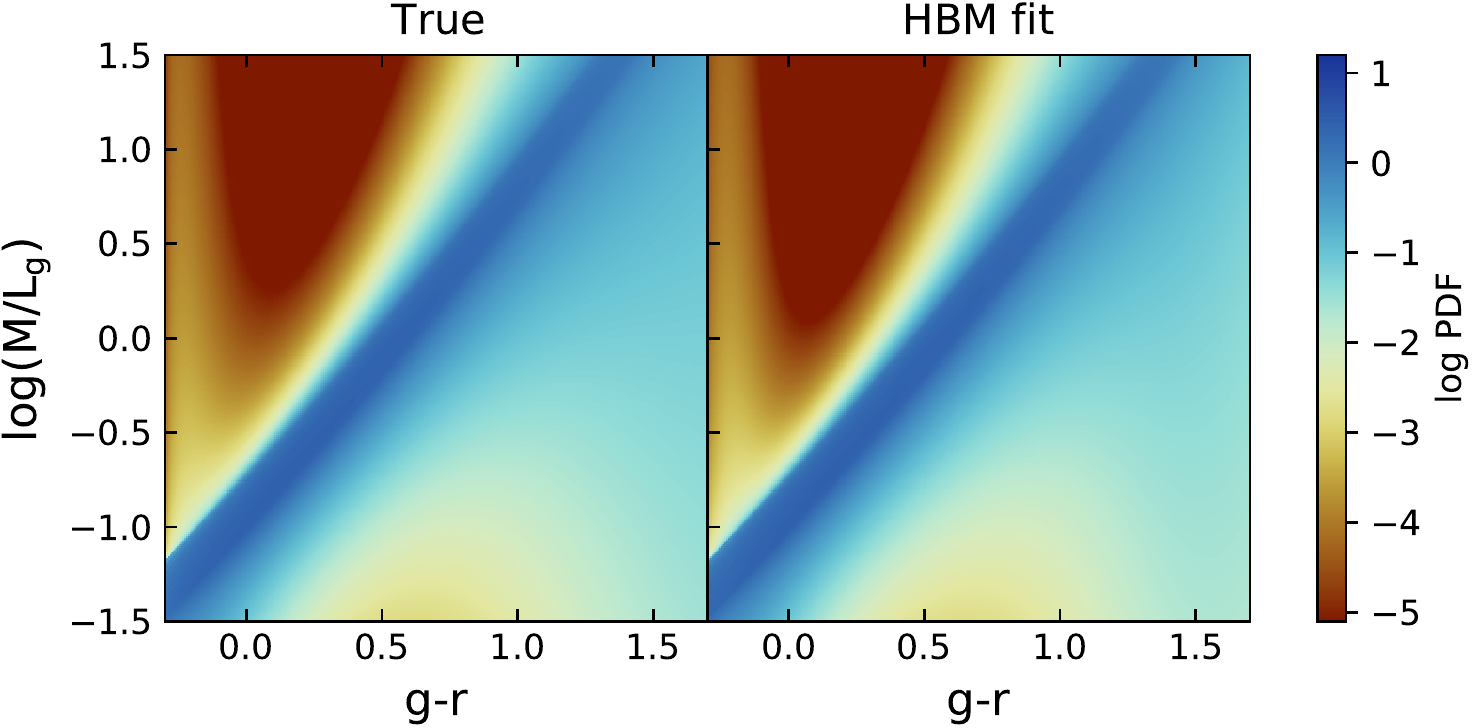}
    \caption{Comparison between the probability density function of the input mock data to the posteriors of their HBM fits. The recovered PDF is similar to the input one to high precision.}
    \label{fig:mock}
\end{figure*}

Figure~\ref{fig:mock} shows that our HBM successfully reproduces the mock data. We compare here the PDF of the true sgt model from which we simulated the mock data to the posterior of HBM in the \gr{}--$\logML$ plane. They are very similar except that the HBM result has less scatter at the blue and red ends.
We find that the likelihood for the best-fit parameter of the HBM fit is slightly higher than the true solution. This implies that we are still sensitive to the number of galaxies and the errors of the \gr{} and $\logML$. Our population model needs more data to recover the exact parameters. Regardless of that, the distribution of $\logML$ can be measured to relatively high precision.

\bibliography{main}{}

\begin{thebibliography}{}
\expandafter\ifx\csname natexlab\endcsname\relax\def\natexlab#1{#1}\fi
\providecommand{\url}[1]{\href{#1}{#1}}
\providecommand{\dodoi}[1]{doi:~\href{http://doi.org/#1}{\nolinkurl{#1}}}
\providecommand{\doeprint}[1]{\href{http://ascl.net/#1}{\nolinkurl{http://ascl.net/#1}}}
\providecommand{\doarXiv}[1]{\href{https://arxiv.org/abs/#1}{\nolinkurl{https://arxiv.org/abs/#1}}}

\bibitem[{{Astropy Collaboration} {et~al.}(2013){Astropy Collaboration},
  {Robitaille}, {Tollerud}, {Greenfield}, {Droettboom}, {Bray}, {Aldcroft},
  {Davis}, {Ginsburg}, {Price-Whelan}, {Kerzendorf}, {Conley}, {Crighton},
  {Barbary}, {Muna}, {Ferguson}, {Grollier}, {Parikh}, {Nair}, {Unther},
  {Deil}, {Woillez}, {Conseil}, {Kramer}, {Turner}, {Singer}, {Fox}, {Weaver},
  {Zabalza}, {Edwards}, {Azalee Bostroem}, {Burke}, {Casey}, {Crawford},
  {Dencheva}, {Ely}, {Jenness}, {Labrie}, {Lim}, {Pierfederici}, {Pontzen},
  {Ptak}, {Refsdal}, {Servillat}, \& {Streicher}}]{astropy:2013}
{Astropy Collaboration}, {Robitaille}, T.~P., {Tollerud}, E.~J., {et~al.} 2013,
  \aap, 558, A33, \dodoi{10.1051/0004-6361/201322068}

\bibitem[{{Astropy Collaboration} {et~al.}(2018){Astropy Collaboration},
  {Price-Whelan}, {Sip{\H{o}}cz}, {G{\"u}nther}, {Lim}, {Crawford}, {Conseil},
  {Shupe}, {Craig}, {Dencheva}, {Ginsburg}, {Vand erPlas}, {Bradley},
  {P{\'e}rez-Su{\'a}rez}, {de Val-Borro}, {Aldcroft}, {Cruz}, {Robitaille},
  {Tollerud}, {Ardelean}, {Babej}, {Bach}, {Bachetti}, {Bakanov}, {Bamford},
  {Barentsen}, {Barmby}, {Baumbach}, {Berry}, {Biscani}, {Boquien}, {Bostroem},
  {Bouma}, {Brammer}, {Bray}, {Breytenbach}, {Buddelmeijer}, {Burke},
  {Calderone}, {Cano Rodr{\'\i}guez}, {Cara}, {Cardoso}, {Cheedella}, {Copin},
  {Corrales}, {Crichton}, {D'Avella}, {Deil}, {Depagne}, {Dietrich}, {Donath},
  {Droettboom}, {Earl}, {Erben}, {Fabbro}, {Ferreira}, {Finethy}, {Fox},
  {Garrison}, {Gibbons}, {Goldstein}, {Gommers}, {Greco}, {Greenfield},
  {Groener}, {Grollier}, {Hagen}, {Hirst}, {Homeier}, {Horton}, {Hosseinzadeh},
  {Hu}, {Hunkeler}, {Ivezi{\'c}}, {Jain}, {Jenness}, {Kanarek}, {Kendrew},
  {Kern}, {Kerzendorf}, {Khvalko}, {King}, {Kirkby}, {Kulkarni}, {Kumar},
  {Lee}, {Lenz}, {Littlefair}, {Ma}, {Macleod}, {Mastropietro}, {McCully},
  {Montagnac}, {Morris}, {Mueller}, {Mumford}, {Muna}, {Murphy}, {Nelson},
  {Nguyen}, {Ninan}, {N{\"o}the}, {Ogaz}, {Oh}, {Parejko}, {Parley}, {Pascual},
  {Patil}, {Patil}, {Plunkett}, {Prochaska}, {Rastogi}, {Reddy Janga},
  {Sabater}, {Sakurikar}, {Seifert}, {Sherbert}, {Sherwood-Taylor}, {Shih},
  {Sick}, {Silbiger}, {Singanamalla}, {Singer}, {Sladen}, {Sooley},
  {Sornarajah}, {Streicher}, {Teuben}, {Thomas}, {Tremblay}, {Turner},
  {Terr{\'o}n}, {van Kerkwijk}, {de la Vega}, {Watkins}, {Weaver}, {Whitmore},
  {Woillez}, {Zabalza}, \& {Astropy Contributors}}]{astropy:2018}
{Astropy Collaboration}, {Price-Whelan}, A.~M., {Sip{\H{o}}cz}, B.~M., {et~al.}
  2018, \aj, 156, 123, \dodoi{10.3847/1538-3881/aabc4f}

\bibitem[{{Bell} \& {de Jong}(2001)}]{Bell&deJong2001}
{Bell}, E.~F., \& {de Jong}, R.~S. 2001, \apj, 550, 212, \dodoi{10.1086/319728}

\bibitem[{{Bell} {et~al.}(2003){Bell}, {McIntosh}, {Katz}, \&
  {Weinberg}}]{Bell2003}
{Bell}, E.~F., {McIntosh}, D.~H., {Katz}, N., \& {Weinberg}, M.~D. 2003, \apjs,
  149, 289, \dodoi{10.1086/378847}

\bibitem[{{Blanton} \& {Roweis}(2007)}]{Blanton2007}
{Blanton}, M.~R., \& {Roweis}, S. 2007, \aj, 133, 734, \dodoi{10.1086/510127}

\bibitem[{{Brammer} {et~al.}(2008){Brammer}, {van Dokkum}, \&
  {Coppi}}]{2008ApJ...686.1503B}
{Brammer}, G.~B., {van Dokkum}, P.~G., \& {Coppi}, P. 2008, \apj, 686, 1503,
  \dodoi{10.1086/591786}

\bibitem[{{Bruzual} \& {Charlot}(2003)}]{BC2003}
{Bruzual}, G., \& {Charlot}, S. 2003, \mnras, 344, 1000,
  \dodoi{10.1046/j.1365-8711.2003.06897.x}

\bibitem[{{Byler} {et~al.}(2017){Byler}, {Dalcanton}, {Conroy}, \&
  {Johnson}}]{Byler2017}
{Byler}, N., {Dalcanton}, J.~J., {Conroy}, C., \& {Johnson}, B.~D. 2017, \apj,
  840, 44, \dodoi{10.3847/1538-4357/aa6c66}

\bibitem[{{Calzetti} {et~al.}(2000){Calzetti}, {Armus}, {Bohlin}, {Kinney},
  {Koornneef}, \& {Storchi-Bergmann}}]{Calzetti2000}
{Calzetti}, D., {Armus}, L., {Bohlin}, R.~C., {et~al.} 2000, \apj, 533, 682,
  \dodoi{10.1086/308692}

\bibitem[{{Cappellari}(2017)}]{PPXF2017}
{Cappellari}, M. 2017, \mnras, 466, 798, \dodoi{10.1093/mnras/stw3020}

\bibitem[{{Carnall} {et~al.}(2019){Carnall}, {Leja}, {Johnson}, {McLure},
  {Dunlop}, \& {Conroy}}]{Carnall2019}
{Carnall}, A.~C., {Leja}, J., {Johnson}, B.~D., {et~al.} 2019, \apj, 873, 44,
  \dodoi{10.3847/1538-4357/ab04a2}

\bibitem[{{Chabrier}(2003)}]{Chabrier2003}
{Chabrier}, G. 2003, \pasp, 115, 763, \dodoi{10.1086/376392}

\bibitem[{{Chevallard} {et~al.}(2013){Chevallard}, {Charlot}, {Wandelt}, \&
  {Wild}}]{Chevallard2013}
{Chevallard}, J., {Charlot}, S., {Wandelt}, B., \& {Wild}, V. 2013, \mnras,
  432, 2061, \dodoi{10.1093/mnras/stt523}

\bibitem[{{Chilingarian} {et~al.}(2010){Chilingarian}, {Melchior}, \&
  {Zolotukhin}}]{Chilingarian2010}
{Chilingarian}, I.~V., {Melchior}, A.-L., \& {Zolotukhin}, I.~Y. 2010, \mnras,
  405, 1409, \dodoi{10.1111/j.1365-2966.2010.16506.x}

\bibitem[{{Choi} {et~al.}(2016){Choi}, {Dotter}, {Conroy}, {Cantiello},
  {Paxton}, \& {Johnson}}]{MIST1}
{Choi}, J., {Dotter}, A., {Conroy}, C., {et~al.} 2016, \apj, 823, 102,
  \dodoi{10.3847/0004-637X/823/2/102}

\bibitem[{{Conroy}(2013)}]{Conroy2013}
{Conroy}, C. 2013, \araa, 51, 393, \dodoi{10.1146/annurev-astro-082812-141017}

\bibitem[{{Conroy} {et~al.}(2009){Conroy}, {Gunn}, \& {White}}]{Conroy2009}
{Conroy}, C., {Gunn}, J.~E., \& {White}, M. 2009, \apj, 699, 486,
  \dodoi{10.1088/0004-637X/699/1/486}

\bibitem[{{Courteau} {et~al.}(2014){Courteau}, {Cappellari}, {de Jong},
  {Dutton}, {Emsellem}, {Hoekstra}, {Koopmans}, {Mamon}, {Maraston}, {Treu}, \&
  {Widrow}}]{Courteau2014}
{Courteau}, S., {Cappellari}, M., {de Jong}, R.~S., {et~al.} 2014, Reviews of
  Modern Physics, 86, 47, \dodoi{10.1103/RevModPhys.86.47}

\bibitem[{{Cowie} {et~al.}(1996){Cowie}, {Songaila}, {Hu}, \&
  {Cohen}}]{Cowie1996}
{Cowie}, L.~L., {Songaila}, A., {Hu}, E.~M., \& {Cohen}, J.~G. 1996, \aj, 112,
  839, \dodoi{10.1086/118058}

\bibitem[{{Crain} {et~al.}(2015){Crain}, {Schaye}, {Bower}, {Furlong},
  {Schaller}, {Theuns}, {Dalla Vecchia}, {Frenk}, {McCarthy}, {Helly},
  {Jenkins}, {Rosas-Guevara}, {White}, \& {Trayford}}]{Crain2015}
{Crain}, R.~A., {Schaye}, J., {Bower}, R.~G., {et~al.} 2015, \mnras, 450, 1937,
  \dodoi{10.1093/mnras/stv725}

\bibitem[{{da Cunha} {et~al.}(2008){da Cunha}, {Charlot}, \&
  {Elbaz}}]{daCunha2008}
{da Cunha}, E., {Charlot}, S., \& {Elbaz}, D. 2008, \mnras, 388, 1595,
  \dodoi{10.1111/j.1365-2966.2008.13535.x}

\bibitem[{{de Graaff} {et~al.}(2021){de Graaff}, {Bezanson}, {Franx}, {van der
  Wel}, {Holden}, {van de Sande}, {Bell}, {D'Eugenio}, {Maseda}, {Muzzin},
  {Sobral}, {Straatman}, \& {Wu}}]{deGraaff2021}
{de Graaff}, A., {Bezanson}, R., {Franx}, M., {et~al.} 2021, \apj, 913, 103,
  \dodoi{10.3847/1538-4357/abf1e7}

\bibitem[{{Dotter}(2016)}]{MIST0}
{Dotter}, A. 2016, \apjs, 222, 8, \dodoi{10.3847/0067-0049/222/1/8}

\bibitem[{{Foreman-Mackey} {et~al.}(2014){Foreman-Mackey}, {Sick}, \&
  {Johnson}}]{python-fsps}
{Foreman-Mackey}, D., {Sick}, J., \& {Johnson}, B. 2014, {python-fsps: Python
  bindings to FSPS (v0.1.1)}, v0.1.1, Zenodo,  Zenodo,
  \dodoi{10.5281/zenodo.12157}

\bibitem[{{Gallazzi} {et~al.}(2005){Gallazzi}, {Charlot}, {Brinchmann},
  {White}, \& {Tremonti}}]{massmet}
{Gallazzi}, A., {Charlot}, S., {Brinchmann}, J., {White}, S. D.~M., \&
  {Tremonti}, C.~A. 2005, \mnras, 362, 41,
  \dodoi{10.1111/j.1365-2966.2005.09321.x}

\bibitem[{{Garc{\'\i}a-Benito} {et~al.}(2019){Garc{\'\i}a-Benito},
  {Gonz{\'a}lez Delgado}, {P{\'e}rez}, {Cid Fernandes}, {S{\'a}nchez}, \& {de
  Amorim}}]{Garcia2019}
{Garc{\'\i}a-Benito}, R., {Gonz{\'a}lez Delgado}, R.~M., {P{\'e}rez}, E.,
  {et~al.} 2019, \aap, 621, A120, \dodoi{10.1051/0004-6361/201833993}

\bibitem[{{Ge} {et~al.}(2021){Ge}, {Mao}, {Lu}, {Cappellari}, {Long}, \&
  {Yan}}]{Ge2021}
{Ge}, J., {Mao}, S., {Lu}, Y., {et~al.} 2021, \mnras, 507, 2488,
  \dodoi{10.1093/mnras/stab2341}

\bibitem[{{Ge} {et~al.}(2019){Ge}, {Mao}, {Lu}, {Cappellari}, \&
  {Yan}}]{Ge2019}
{Ge}, J., {Mao}, S., {Lu}, Y., {Cappellari}, M., \& {Yan}, R. 2019, \mnras,
  485, 1675, \dodoi{10.1093/mnras/stz418}

\bibitem[{Harris {et~al.}(2020)Harris, Millman, van~der Walt, Gommers,
  Virtanen, Cournapeau, Wieser, Taylor, Berg, Smith, Kern, Picus, Hoyer, van
  Kerkwijk, Brett, Haldane, del R{\'{i}}o, Wiebe, Peterson,
  G{\'{e}}rard-Marchant, Sheppard, Reddy, Weckesser, Abbasi, Gohlke, \&
  Oliphant}]{harris2020array}
Harris, C.~R., Millman, K.~J., van~der Walt, S.~J., {et~al.} 2020, Nature, 585,
  357, \dodoi{10.1038/s41586-020-2649-2}

\bibitem[{{Hogg} {et~al.}(2002){Hogg}, {Baldry}, {Blanton}, \&
  {Eisenstein}}]{Hogg2002}
{Hogg}, D.~W., {Baldry}, I.~K., {Blanton}, M.~R., \& {Eisenstein}, D.~J. 2002,
  arXiv e-prints, astro.
\newblock \doarXiv{astro-ph/0210394}

\bibitem[{Hunter(2007)}]{Hunter2007}
Hunter, J.~D. 2007, Computing in Science \& Engineering, 9, 90,
  \dodoi{10.1109/MCSE.2007.55}

\bibitem[{{Into} \& {Portinari}(2013)}]{Into2013}
{Into}, T., \& {Portinari}, L. 2013, \mnras, 430, 2715,
  \dodoi{10.1093/mnras/stt071}

\bibitem[{{Iyer} {et~al.}(2020){Iyer}, {Tacchella}, {Genel}, {Hayward},
  {Hernquist}, {Brooks}, {Caplar}, {Dav{\'e}}, {Diemer}, {Forbes}, {Gawiser},
  {Somerville}, \& {Starkenburg}}]{Iyer2020}
{Iyer}, K.~G., {Tacchella}, S., {Genel}, S., {et~al.} 2020, \mnras, 498, 430,
  \dodoi{10.1093/mnras/staa2150}

\bibitem[{{Johnson} {et~al.}(2021){Johnson}, {Leja}, {Conroy}, \&
  {Speagle}}]{Prospector-2}
{Johnson}, B.~D., {Leja}, J., {Conroy}, C., \& {Speagle}, J.~S. 2021, \apjs,
  254, 22, \dodoi{10.3847/1538-4365/abef67}

\bibitem[{{Kelly}(2007)}]{Kelly2007}
{Kelly}, B.~C. 2007, \apj, 665, 1489, \dodoi{10.1086/519947}

\bibitem[{{Kriek} {et~al.}(2009){Kriek}, {van Dokkum}, {Labb{\'e}}, {Franx},
  {Illingworth}, {Marchesini}, \& {Quadri}}]{2009ApJ...700..221K}
{Kriek}, M., {van Dokkum}, P.~G., {Labb{\'e}}, I., {et~al.} 2009, \apj, 700,
  221, \dodoi{10.1088/0004-637X/700/1/221}

\bibitem[{{Kriek} {et~al.}(2010){Kriek}, {Labb{\'e}}, {Conroy}, {Whitaker},
  {van Dokkum}, {Brammer}, {Franx}, {Illingworth}, {Marchesini}, {Muzzin},
  {Quadri}, \& {Rudnick}}]{Kriek2010}
{Kriek}, M., {Labb{\'e}}, I., {Conroy}, C., {et~al.} 2010, \apjl, 722, L64,
  \dodoi{10.1088/2041-8205/722/1/L64}

\bibitem[{{Leja} {et~al.}(2019{\natexlab{a}}){Leja}, {Carnall}, {Johnson},
  {Conroy}, \& {Speagle}}]{Leja2019a}
{Leja}, J., {Carnall}, A.~C., {Johnson}, B.~D., {Conroy}, C., \& {Speagle},
  J.~S. 2019{\natexlab{a}}, \apj, 876, 3, \dodoi{10.3847/1538-4357/ab133c}

\bibitem[{{Leja} {et~al.}(2017){Leja}, {Johnson}, {Conroy}, {van Dokkum}, \&
  {Byler}}]{Prospector-1}
{Leja}, J., {Johnson}, B.~D., {Conroy}, C., {van Dokkum}, P.~G., \& {Byler}, N.
  2017, \apj, 837, 170, \dodoi{10.3847/1538-4357/aa5ffe}

\bibitem[{{Leja} {et~al.}(2020){Leja}, {Speagle}, {Johnson}, {Conroy}, {van
  Dokkum}, \& {Franx}}]{prospector-massfunc}
{Leja}, J., {Speagle}, J.~S., {Johnson}, B.~D., {et~al.} 2020, \apj, 893, 111,
  \dodoi{10.3847/1538-4357/ab7e27}

\bibitem[{{Leja} {et~al.}(2019{\natexlab{b}}){Leja}, {Johnson}, {Conroy}, {van
  Dokkum}, {Speagle}, {Brammer}, {Momcheva}, {Skelton}, {Whitaker}, {Franx}, \&
  {Nelson}}]{2019ApJ...877..140L}
{Leja}, J., {Johnson}, B.~D., {Conroy}, C., {et~al.} 2019{\natexlab{b}}, \apj,
  877, 140, \dodoi{10.3847/1538-4357/ab1d5a}

\bibitem[{{L{\'o}pez-Sanjuan} {et~al.}(2019){L{\'o}pez-Sanjuan},
  {D{\'\i}az-Garc{\'\i}a}, {Cenarro}, {Fern{\'a}ndez-Soto}, {Viironen},
  {Molino}, {Ben{\'\i}tez}, {Crist{\'o}bal-Hornillos}, {Moles}, {Varela},
  {Arnalte-Mur}, {Ascaso}, {Castander}, {Cervi{\~n}o}, {Gonz{\'a}lez Delgado},
  {Husillos}, {M{\'a}rquez}, {Masegosa}, {Del Olmo}, {Povi{\'c}}, \&
  {Perea}}]{ALHAMBRA2019}
{L{\'o}pez-Sanjuan}, C., {D{\'\i}az-Garc{\'\i}a}, L.~A., {Cenarro}, A.~J.,
  {et~al.} 2019, \aap, 622, A51, \dodoi{10.1051/0004-6361/201833402}

\bibitem[{{Loredo} \& {Hendry}(2019)}]{Loredo&Hendry2019}
{Loredo}, T.~J., \& {Hendry}, M.~A. 2019, arXiv e-prints, arXiv:1911.12337.
\newblock \doarXiv{1911.12337}

\bibitem[{{Lower} {et~al.}(2020){Lower}, {Narayanan}, {Leja}, {Johnson},
  {Conroy}, \& {Dav{\'e}}}]{Lower2020}
{Lower}, S., {Narayanan}, D., {Leja}, J., {et~al.} 2020, \apj, 904, 33,
  \dodoi{10.3847/1538-4357/abbfa7}

\bibitem[{{Ma{\l}ek} {et~al.}(2018){Ma{\l}ek}, {Buat}, {Roehlly}, {Burgarella},
  {Hurley}, {Shirley}, {Duncan}, {Efstathiou}, {Papadopoulos}, {Vaccari},
  {Farrah}, {Marchetti}, \& {Oliver}}]{Malek2018}
{Ma{\l}ek}, K., {Buat}, V., {Roehlly}, Y., {et~al.} 2018, \aap, 620, A50,
  \dodoi{10.1051/0004-6361/201833131}

\bibitem[{{Maraston} {et~al.}(2010){Maraston}, {Pforr}, {Renzini}, {Daddi},
  {Dickinson}, {Cimatti}, \& {Tonini}}]{Maraston2010}
{Maraston}, C., {Pforr}, J., {Renzini}, A., {et~al.} 2010, \mnras, 407, 830,
  \dodoi{10.1111/j.1365-2966.2010.16973.x}

\bibitem[{{Marchesini} {et~al.}(2009){Marchesini}, {van Dokkum}, {F{\"o}rster
  Schreiber}, {Franx}, {Labb{\'e}}, \& {Wuyts}}]{Marchesini2009}
{Marchesini}, D., {van Dokkum}, P.~G., {F{\"o}rster Schreiber}, N.~M., {et~al.}
  2009, \apj, 701, 1765, \dodoi{10.1088/0004-637X/701/2/1765}

\bibitem[{{Maseda} {et~al.}(2018){Maseda}, {van der Wel}, {Rix}, {Momcheva},
  {Brammer}, {Franx}, {Lundgren}, {Skelton}, \& {Whitaker}}]{Maseda2018}
{Maseda}, M.~V., {van der Wel}, A., {Rix}, H.-W., {et~al.} 2018, \apj, 854, 29,
  \dodoi{10.3847/1538-4357/aaa76e}

\bibitem[{{McGaugh} \& {Schombert}(2014)}]{McGaugh2014}
{McGaugh}, S.~S., \& {Schombert}, J.~M. 2014, \aj, 148, 77,
  \dodoi{10.1088/0004-6256/148/5/77}

\bibitem[{{Mitchell} {et~al.}(2013){Mitchell}, {Lacey}, {Baugh}, \&
  {Cole}}]{Mitchell2013}
{Mitchell}, P.~D., {Lacey}, C.~G., {Baugh}, C.~M., \& {Cole}, S. 2013, \mnras,
  435, 87, \dodoi{10.1093/mnras/stt1280}

\bibitem[{{Momcheva} {et~al.}(2016){Momcheva}, {Brammer}, {van Dokkum},
  {Skelton}, {Whitaker}, {Nelson}, {Fumagalli}, {Maseda}, {Leja}, {Franx},
  {Rix}, {Bezanson}, {Da Cunha}, {Dickey}, {F{\"o}rster Schreiber},
  {Illingworth}, {Kriek}, {Labb{\'e}}, {Ulf Lange}, {Lundgren}, {Magee},
  {Marchesini}, {Oesch}, {Pacifici}, {Patel}, {Price}, {Tal}, {Wake}, {van der
  Wel}, \& {Wuyts}}]{2016ApJS..225...27M}
{Momcheva}, I.~G., {Brammer}, G.~B., {van Dokkum}, P.~G., {et~al.} 2016, \apjs,
  225, 27, \dodoi{10.3847/0067-0049/225/2/27}

\bibitem[{{Muzzin} {et~al.}(2013){Muzzin}, {Marchesini}, {Stefanon}, {Franx},
  {Milvang-Jensen}, {Dunlop}, {Fynbo}, {Brammer}, {Labb{\'e}}, \& {van
  Dokkum}}]{Muzzin2013ApJS}
{Muzzin}, A., {Marchesini}, D., {Stefanon}, M., {et~al.} 2013, \apjs, 206, 8,
  \dodoi{10.1088/0067-0049/206/1/8}

\bibitem[{{Nagaraj} {et~al.}(2022){Nagaraj}, {Forbes}, {Leja},
  {Foreman-Mackey}, \& {Hayward}}]{Nagaraj2022}
{Nagaraj}, G., {Forbes}, J.~C., {Leja}, J., {Foreman-Mackey}, D., \& {Hayward},
  C.~C. 2022, arXiv e-prints, arXiv:2202.05102.
\newblock \doarXiv{2202.05102}

\bibitem[{{Nguyen} {et~al.}(2020){Nguyen}, {den Brok}, {Seth}, {Davis},
  {Greene}, {Cappellari}, {Jensen}, {Thater}, {Iguchi}, {Imanishi}, {Izumi},
  {Nyland}, {Neumayer}, {Nakanishi}, {Nguyen}, {Tsukui}, {Bureau}, {Onishi},
  {Quang}, \& {Le}}]{Nguyen2020}
{Nguyen}, D.~D., {den Brok}, M., {Seth}, A.~C., {et~al.} 2020, \apj, 892, 68,
  \dodoi{10.3847/1538-4357/ab77aa}

\bibitem[{{Noll} {et~al.}(2009){Noll}, {Burgarella}, {Giovannoli}, {Buat},
  {Marcillac}, \& {Mu{\~n}oz-Mateos}}]{Noll2009}
{Noll}, S., {Burgarella}, D., {Giovannoli}, E., {et~al.} 2009, \aap, 507, 1793,
  \dodoi{10.1051/0004-6361/200912497}

\bibitem[{{Pacifici} {et~al.}(2013){Pacifici}, {Kassin}, {Weiner}, {Charlot},
  \& {Gardner}}]{Pacifici2013}
{Pacifici}, C., {Kassin}, S.~A., {Weiner}, B., {Charlot}, S., \& {Gardner},
  J.~P. 2013, \apjl, 762, L15, \dodoi{10.1088/2041-8205/762/1/L15}

\bibitem[{{Papovich} {et~al.}(2001){Papovich}, {Dickinson}, \&
  {Ferguson}}]{Papovich2001}
{Papovich}, C., {Dickinson}, M., \& {Ferguson}, H.~C. 2001, \apj, 559, 620,
  \dodoi{10.1086/322412}

\bibitem[{Pedregosa {et~al.}(2011)Pedregosa, Varoquaux, Gramfort, Michel,
  Thirion, Grisel, Blondel, Prettenhofer, Weiss, Dubourg, Vanderplas, Passos,
  Cournapeau, Brucher, Perrot, \& Duchesnay}]{Pedregosa2011scikit}
Pedregosa, F., Varoquaux, G., Gramfort, A., {et~al.} 2011, Journal of Machine
  Learning Research, 12, 2825

\bibitem[{{Pforr} {et~al.}(2012){Pforr}, {Maraston}, \& {Tonini}}]{Pforr2012}
{Pforr}, J., {Maraston}, C., \& {Tonini}, C. 2012, \mnras, 422, 3285,
  \dodoi{10.1111/j.1365-2966.2012.20848.x}

\bibitem[{{Portinari} {et~al.}(2004){Portinari}, {Sommer-Larsen}, \&
  {Tantalo}}]{Portinari2004}
{Portinari}, L., {Sommer-Larsen}, J., \& {Tantalo}, R. 2004, \mnras, 347, 691,
  \dodoi{10.1111/j.1365-2966.2004.07207.x}

\bibitem[{{Salim} \& {Narayanan}(2020)}]{dustlaw2020review}
{Salim}, S., \& {Narayanan}, D. 2020, \araa, 58, 529,
  \dodoi{10.1146/annurev-astro-032620-021933}

\bibitem[{{Salmon} {et~al.}(2016){Salmon}, {Papovich}, {Long}, {Willner},
  {Finkelstein}, {Ferguson}, {Dickinson}, {Duncan}, {Faber}, {Hathi},
  {Koekemoer}, {Kurczynski}, {Newman}, {Pacifici}, {P{\'e}rez-Gonz{\'a}lez}, \&
  {Pforr}}]{Salmon2016}
{Salmon}, B., {Papovich}, C., {Long}, J., {et~al.} 2016, \apj, 827, 20,
  \dodoi{10.3847/0004-637X/827/1/20}

\bibitem[{{S{\'a}nchez-Bl{\'a}zquez} {et~al.}(2006){S{\'a}nchez-Bl{\'a}zquez},
  {Peletier}, {Jim{\'e}nez-Vicente}, {Cardiel}, {Cenarro},
  {Falc{\'o}n-Barroso}, {Gorgas}, {Selam}, \& {Vazdekis}}]{MILES}
{S{\'a}nchez-Bl{\'a}zquez}, P., {Peletier}, R.~F., {Jim{\'e}nez-Vicente}, J.,
  {et~al.} 2006, \mnras, 371, 703, \dodoi{10.1111/j.1365-2966.2006.10699.x}

\bibitem[{{Senchyna} {et~al.}(2019){Senchyna}, {Stark}, {Chevallard},
  {Charlot}, {Jones}, \& {Vidal-Garc{\'\i}a}}]{Senchyna2019}
{Senchyna}, P., {Stark}, D.~P., {Chevallard}, J., {et~al.} 2019, \mnras, 488,
  3492, \dodoi{10.1093/mnras/stz1907}

\bibitem[{{Shapley} {et~al.}(2001){Shapley}, {Steidel}, {Adelberger},
  {Dickinson}, {Giavalisco}, \& {Pettini}}]{Shapley2001}
{Shapley}, A.~E., {Steidel}, C.~C., {Adelberger}, K.~L., {et~al.} 2001, \apj,
  562, 95, \dodoi{10.1086/323432}

\bibitem[{{Skelton} {et~al.}(2014){Skelton}, {Whitaker}, {Momcheva}, {Brammer},
  {van Dokkum}, {Labb{\'e}}, {Franx}, {van der Wel}, {Bezanson}, {Da Cunha},
  {Fumagalli}, {F{\"o}rster Schreiber}, {Kriek}, {Leja}, {Lundgren}, {Magee},
  {Marchesini}, {Maseda}, {Nelson}, {Oesch}, {Pacifici}, {Patel}, {Price},
  {Rix}, {Tal}, {Wake}, \& {Wuyts}}]{Skelton2014}
{Skelton}, R.~E., {Whitaker}, K.~E., {Momcheva}, I.~G., {et~al.} 2014, \apjs,
  214, 24, \dodoi{10.1088/0067-0049/214/2/24}

\bibitem[{{Somerville} \& {Dav{\'e}}(2015)}]{Somerville&Dave2015}
{Somerville}, R.~S., \& {Dav{\'e}}, R. 2015, \araa, 53, 51,
  \dodoi{10.1146/annurev-astro-082812-140951}

\bibitem[{{Speagle}(2020)}]{dynesty}
{Speagle}, J.~S. 2020, \mnras, 493, 3132, \dodoi{10.1093/mnras/staa278}

\bibitem[{{Szomoru} {et~al.}(2013){Szomoru}, {Franx}, {van Dokkum}, {Trenti},
  {Illingworth}, {Labb{\'e}}, \& {Oesch}}]{Szomoru2013}
{Szomoru}, D., {Franx}, M., {van Dokkum}, P.~G., {et~al.} 2013, \apj, 763, 73,
  \dodoi{10.1088/0004-637X/763/2/73}

\bibitem[{{Tacchella} {et~al.}(2022){Tacchella}, {Conroy}, {Faber}, {Johnson},
  {Leja}, {Barro}, {Cunningham}, {Deason}, {Guhathakurta}, {Guo}, {Hernquist},
  {Koo}, {McKinnon}, {Rockosi}, {Speagle}, {van Dokkum}, \&
  {Yesuf}}]{Tacchella2022}
{Tacchella}, S., {Conroy}, C., {Faber}, S.~M., {et~al.} 2022, \apj, 926, 134,
  \dodoi{10.3847/1538-4357/ac449b}

\bibitem[{{Taylor} {et~al.}(2010){Taylor}, {Franx}, {Brinchmann}, {van der
  Wel}, \& {van Dokkum}}]{Taylor2010}
{Taylor}, E.~N., {Franx}, M., {Brinchmann}, J., {van der Wel}, A., \& {van
  Dokkum}, P.~G. 2010, \apj, 722, 1, \dodoi{10.1088/0004-637X/722/1/1}

\bibitem[{{Taylor} {et~al.}(2011){Taylor}, {Hopkins}, {Baldry}, {Brown},
  {Driver}, {Kelvin}, {Hill}, {Robotham}, {Bland-Hawthorn}, {Jones}, {Sharp},
  {Thomas}, {Liske}, {Loveday}, {Norberg}, {Peacock}, {Bamford}, {Brough},
  {Colless}, {Cameron}, {Conselice}, {Croom}, {Frenk}, {Gunawardhana},
  {Kuijken}, {Nichol}, {Parkinson}, {Phillipps}, {Pimbblet}, {Popescu},
  {Prescott}, {Sutherland}, {Tuffs}, {van Kampen}, \&
  {Wijesinghe}}]{Taylor2011GAMA}
{Taylor}, E.~N., {Hopkins}, A.~M., {Baldry}, I.~K., {et~al.} 2011, \mnras, 418,
  1587, \dodoi{10.1111/j.1365-2966.2011.19536.x}

\bibitem[{{van de Sande} {et~al.}(2015){van de Sande}, {Kriek}, {Franx},
  {Bezanson}, \& {van Dokkum}}]{van-de-Sande2015}
{van de Sande}, J., {Kriek}, M., {Franx}, M., {Bezanson}, R., \& {van Dokkum},
  P.~G. 2015, \apj, 799, 125, \dodoi{10.1088/0004-637X/799/2/125}

\bibitem[{Virtanen {et~al.}(2020)Virtanen, Gommers, Oliphant, Haberland, Reddy,
  Cournapeau, Burovski, Peterson, Weckesser, Bright, {van der Walt}, Brett,
  Wilson, Millman, Mayorov, Nelson, Jones, Kern, Larson, Carey, Polat, Feng,
  Moore, {VanderPlas}, Laxalde, Perktold, Cimrman, Henriksen, Quintero, Harris,
  Archibald, Ribeiro, Pedregosa, {van Mulbregt}, \& {SciPy 1.0
  Contributors}}]{2020SciPy-NMeth}
Virtanen, P., Gommers, R., Oliphant, T.~E., {et~al.} 2020, Nature Methods, 17,
  261, \dodoi{10.1038/s41592-019-0686-2}

\bibitem[{{Walcher} {et~al.}(2011){Walcher}, {Groves}, {Budav{\'a}ri}, \&
  {Dale}}]{Walcher2011}
{Walcher}, J., {Groves}, B., {Budav{\'a}ri}, T., \& {Dale}, D. 2011, \apss,
  331, 1, \dodoi{10.1007/s10509-010-0458-z}

\bibitem[{{Whitaker} {et~al.}(2014){Whitaker}, {Franx}, {Leja}, {van Dokkum},
  {Henry}, {Skelton}, {Fumagalli}, {Momcheva}, {Brammer}, {Labb{\'e}},
  {Nelson}, \& {Rigby}}]{2014ApJ...795..104W}
{Whitaker}, K.~E., {Franx}, M., {Leja}, J., {et~al.} 2014, \apj, 795, 104,
  \dodoi{10.1088/0004-637X/795/2/104}

\bibitem[{{Willmer}(2018)}]{Willmer2018solarMags}
{Willmer}, C. N.~A. 2018, \apjs, 236, 47, \dodoi{10.3847/1538-4365/aabfdf}

\bibitem[{{Wuyts} {et~al.}(2011){Wuyts}, {F{\"o}rster Schreiber}, {Lutz},
  {Nordon}, {Berta}, {Altieri}, {Andreani}, {Aussel}, {Bongiovanni}, {Cepa},
  {Cimatti}, {Daddi}, {Elbaz}, {Genzel}, {Koekemoer}, {Magnelli}, {Maiolino},
  {McGrath}, {P{\'e}rez Garc{\'\i}a}, {Poglitsch}, {Popesso}, {Pozzi},
  {Sanchez-Portal}, {Sturm}, {Tacconi}, \& {Valtchanov}}]{Wuyts2011}
{Wuyts}, S., {F{\"o}rster Schreiber}, N.~M., {Lutz}, D., {et~al.} 2011, \apj,
  738, 106, \dodoi{10.1088/0004-637X/738/1/106}

\bibitem[{{Zibetti} {et~al.}(2009){Zibetti}, {Charlot}, \& {Rix}}]{Zibetti2009}
{Zibetti}, S., {Charlot}, S., \& {Rix}, H.-W. 2009, \mnras, 400, 1181,
  \dodoi{10.1111/j.1365-2966.2009.15528.x}

\end{thebibliography}
\bibliographystyle{aasjournal}

\end{CJK*}
\end{document}